\documentclass[12pt,fleqn]{article}
\usepackage{amssymb,amsmath}
\usepackage[dvips]{graphicx,psfrag}
\begin{document}
\begin{flushright}
KEK-TH-707 \\
{\tt hep-ph/0008269 }\\
August, 2000 \\
\end{flushright}
\vspace*{2cm}
\begin{center}
{\baselineskip 25pt
\large{\bf 
Large Angle MSW Solution in
Grand Unified Theories\\ \vspace{.2cm}
with SU(3) $\times$ U(1) Horizontal Symmetry
}}

\vspace{1cm}

{\large Ryuichiro Kitano\footnote
{email: {\tt ryuichiro.kitano@kek.jp}}$^{* \dagger}$
and Yukihiro Mimura\footnote
{email: {\tt mimura@ccthmail.kek.jp};
{Address after Sep.\ 1, 2000: 
Department of Physics,
Oklahoma State University,
Stillwater, OK, 74078, USA}
}$^*$
}
\vspace{.5cm}

{\small {\it $^*$Theory Group, KEK, Oho 1-1, Tsukuba, Ibaraki 305-0801,Japan \\
and \\
$^\dagger$Department of Particle and Nuclear Physics,
The Graduate University for Advanced Studies,\\
Oho 1-1, Tsukuba, Ibaraki 305-0801, Japan}}

\vspace{.5cm}

\vspace{1.5cm}
{\bf Abstract}
\end{center}

\bigskip
We construct a model with a SU(3) $\times$ U(1)
horizontal symmetry in the context of Grand Unified Theories.
In our models,
the bi-maximal lepton mixing and
suitable neutrino masses for the large angle MSW solution
are obtained without any fine-tuning
due to the spontaneously broken ${\rm SU(3)_H}$ symmetry.
The three generations of 
quarks and leptons are
unified as members of the
${\rm SU(3)_H}$ fundamental representation,
and
the ${\rm U(1)_H}$ charge gives the origin of
the fermion mass hierarchy
and mixing angles.
We present two explicit examples of 
SU(5)$_{\rm GUT}$ and SO(10)$_{\rm GUT}$ models,
in which the Yukawa structures are given successfully.

\newpage

\section{Introduction}
\baselineskip 20pt
Flavor physics is a recent topics in particle physics.
The Cabibbo-Kobayashi-Maskawa (CKM) \cite{CKM} parameters will be measured precisely
in B-factory experiments \cite{Groom:2000in}.
The recent SuperKamiokande data suggest that
the mixing angles of the lepton sector are large
in contrast to those in the quark sector \cite{Fukuda:1998mi}.
For atmospheric neutrinos,
$\nu_\mu \to \nu_\tau$ oscillations are favored and 
the best fit values are $\sin^2 2 \theta = 1.0$ and 
$\Delta m^2 = 3.5 \times 10^{-3}$ eV$^2$ \cite{Scholberg:1999ar}.
For the solar neutrino problem,
the only solution which is not disfavored at 95\% confidence level
is the large angle MSW solution \cite{Wolfenstein:1978ue}
i.e.\ $\sin^2 2 \theta \sim 1$ and 
$\Delta m^2 \sim 10^{-5}-10^{-4}$ eV$^2$ \cite{ysuzuki}.
If we assume three flavor neutrino mixings,
the solar neutrino deficit is explained by
$\nu_e \to \nu_\mu$ oscillations
and the CHOOZ experiment gives a severe constraint on 
the mixing angle between the electron neutrino
and the heaviest neutrino,
$\sin^2 2\theta_{e3} \lesssim 0.1$ \cite{Apollonio:1999ae}.
These two nearly maximal mixings and
a small mixing suggest that
the lepton mixing matrix is the so-called
bi-maximal type, which is 
a completely different structure
to that of the CKM matrix.
Understanding of the patterns of the mixing and fermion masses
is one of the most challenging puzzles for particle physicists.

One approach to answer this puzzle is 
to consider a horizontal 
flavor symmetry \cite{Froggatt:1979nt,Leurer:1994gy,Wilczek:1979xi}.
The horizontal symmetry provides 
the Yukawa coupling structure and enable us
to predict the structure definitely.

The two generation model successfully explains the magnitude of 
the Cabibbo angle \cite{Fritzsch}, namely
\begin{equation}
\sin \theta_{\rm C} \sim \sqrt{\frac{m_d}{m_s}} \ .
\end{equation}
We reproduce this relation by using a flavor symmetry.
However the simple extension to three generations
does not succeed.
For example, the CKM matrix element $V_{cb}$
is not the same order of $\sqrt{m_s/m_b}$ but
instead of order $m_s/m_b$ \cite{Branco}.

The observation of the large mixing between $\nu_\mu-\nu_\tau$
plays an important role in understanding the 2-3 structure.
The structure of the second 
and third generations are revealed to be non-trivial.
The non-parallel family structure has been suggested \cite{Sato:1998hv},
and this structure is also called 
lopsided family structure \cite{Albright:1999jv}
and can be
organized by an Abelian symmetry \cite{Leurer:1994gy}.

The lopsided family structure may give us a neutrino mass matrix
which leads to large angle MSW solution for solar neutrino problem.
However the lopsided family structure does not always predict 
the large mixing angle
unless the mass (squared) ratio is assumed to take a particular value.
There is no reason to obtain large MSW solution
in the lopsided structure.

We will give a simple lopsided family structure
by imposing a SU(3)$_{\rm H} \times$U(1)$_{\rm H}$ horizontal symmetry.
In our model, the fermion mass hierarchy is 
produced by the U(1)$_{\rm H}$ symmetry.
Contrary to the standard approach,
we do not suppose that the vacuum expectation values (VEVs)
of fundamental (anti-fundamental)
representation of SU(3)$_{\rm H}$ symmetry
are non-hierarchical, and this feature provides
large mixing angles in lepton sector.

In this paper,
we assume a Supersymmetric
Grand Unified Theory (SUSY GUT) \cite{GUT} realized
at the high energy scale ($M_G \sim 10^{16}$ GeV).
The SUSY GUT is an attractive theory
which explains many problems in the standard model.
In particular,
the tiny neutrino masses are elegantly explained
by the seesaw mechanism \cite{seesaw} which 
requires heavy right-handed neutrinos with masses
$M_N \sim 10^{14}$ GeV.
This is just below the GUT scale.
It is natural that the origin of the scale $M_N$
is GUT physics, so that
it is important that we understand
the flavor structures in the context of GUTs.

This paper is organized as follows.
In section 2,
we give the Yukawa structures
which are realized in the model
and show that they reproduce
the correct mass ratios and mixing angles for
the quarks and leptons.
In section 3,
we give an example of the SU(5) GUT model
and explain how we can obtain the Yukawa matrices in section 3.
In section 4,
we construct the SO(10) GUT model in which
the large MSW type mass structure is given.
Section 5 is devoted to our conclusions.

\section{Structure of the Model}
\baselineskip 20pt
We will construct the model to explain 
the CKM matrix, fermion masses, 
and lepton mixings suitable for the large angle MSW solution.
by means of a horizontal symmetry.

The observed CKM matrix,
the quark mass ratios 
and lepton mass ratios at GUT scale \cite{Fusaoka} are
approximately 
expressed by powers of 
the Cabibbo angle $\lambda \sim 0.22$
as follows.
\begin{eqnarray}
 V_{\rm CKM} \sim
\left(
\begin{array}{ccc}
 1&\lambda &\lambda^3 \\
 -\lambda&1 &\lambda^2 \\
 -\lambda^3&-\lambda^2 &1 \\
\end{array}
\right)\ , 
\nonumber 
\end{eqnarray}
\begin{eqnarray}
m_u:m_c:m_t \sim \lambda^7 : \lambda^4 : 1 \ ,\ 
m_d:m_s:m_b \sim \lambda^4 : \lambda^2 : 1 \ ,
\label{1}
\end{eqnarray}
\begin{eqnarray}
m_e:m_\mu:m_\tau \sim \lambda^5 :\lambda^2 : 1\ .
\nonumber 
\end{eqnarray}
For the neutrinos,
the mass ratio is given by
\begin{eqnarray}
 m_{\nu_2} : m_{\nu_3} \sim \lambda^{1-2} : 1 \ .
\label{3}
\end{eqnarray}
We obtain eq.(\ref{3}) from 
the $\Delta m^2$ ratio of 
the atmospheric neutrino 
and large MSW solution, which is
\begin{eqnarray}
 \frac{\Delta m^2_{12}}{\Delta m^2_{23}} \sim 10^{-2} 
\sim \lambda^3 \ .
\label{4}
\end{eqnarray}
The Maki-Nakagawa-Sakata (MNS) matrix \cite{Maki:1962mu}
is nearly bi-maximal,
\begin{eqnarray}
 V_{\rm MNS} \sim 
\left(
\begin{array}{ccc}
 1/\sqrt{2} & -1/\sqrt{2} & \epsilon \\
 1/2       & 1/2        & -1/\sqrt{2} \\
 1/2       & 1/2        &  1/\sqrt{2} \\
\end{array}
\right)\ ,
\label{5}
\end{eqnarray}
where the element $\epsilon$ is a small parameter
constrained by the CHOOZ experiment.

We construct a supersymmetric
SU(5) GUT model with a ${\rm SU(3)_H}$ $\times {\rm U(1)_H}$ 
horizontal symmetry whose spontaneous breaking induces
the generation mixing and the mass differences in 
eq.(\ref{1}), (\ref{3}), and (\ref{5}).
The quarks and leptons are unified as
${\rm SU(3)_H}$ triplets,
which explains why there are three generations.

In our model, the Yukawa couplings 
may be expressed 
in the following form:
\begin{equation}
\sum_{i,j=1}^{3} \left(\frac{\Phi}{M_{\rm Pl}}\right)^{x_i+x_j} 
\xi_i \xi_j^{\rm T} \frac1{M_*^2}\ .
\end{equation}
The three $\xi$'s\footnote{
In our models,
we prepare three pairs of $\xi$'s,
though minimal pairs to break ${\rm SU(3)_H}$
is two.
Of course,
in absence of $\xi_3$,
the exterior product $\bar{\xi}_1 \times \bar{\xi}_2$
plays a role of $\xi_3$ effectively.
However, such a minimal choice has
a difficulty in constructing realistic models.
}
are SU(3)$_{\rm H}$ anti-fundamental representations whose
VEVs break the horizontal symmetry at the scale $M_*$.
The $\Phi$ is SU(3)$_{\rm H}$ singlet field, 
and has non-vanishing U(1)$_{\rm H}$ charge.
The VEV of $\Phi$ is $\lambda M_{\rm Pl}$ and provides
a hierarchy in the Yukawa coupling as follows.

With the appropriate ${\rm U(1)_H}$ charge assignment of $\xi$ fields
as shown later in section 3 and 4,
the Yukawa coupling of up-type quark is given by
\begin{equation}
Y_u \sim \left( 
\begin{array}{ccc} 
\langle \xi_3 \rangle & \langle \xi_2 \rangle  & \langle \xi_1 \rangle 
\end{array} \right) 
\left( \begin{array}{ccc}  \lambda^6 & \lambda^5 & \lambda^3 \\
                           \lambda^5 & \lambda^4 & \lambda^2 \\
                           \lambda^3 & \lambda^2 & 1 
       \end{array}
\right)
\left( \begin{array}{c}  \langle \xi_3 \rangle^T \\ 
                         \langle \xi_2 \rangle^T \\
                         \langle \xi_1 \rangle^T 
       \end{array}
\right)/M_*^2 \ .
\label{7}
\end{equation}
The VEVs of $\xi$'s
are given without loss of generality by
\begin{eqnarray}
 \langle \xi_1 \rangle = M_*
\left(
\begin{array}{c}
 0\\
 0\\
 a\\
\end{array}
\right)\ ,\ 
 \langle \xi_2 \rangle = M_*
\left(
\begin{array}{c}
 0\\
 b\\
 c\\
\end{array}
\right)\ ,\ 
 \langle \xi_3 \rangle = M_*
\left(
\begin{array}{c}
 d\\
 e\\
 f\\
\end{array}
\right)\ ,
\label{12}
\end{eqnarray}
where the $a \to f$ are parameters of the order unity.
In this basis,
$Y_u$ is given by
\begin{eqnarray}
 Y_u \sim 
\left(
\begin{array}{ccc}
 \lambda^6 & \lambda^5 & \lambda^3 \\
 \lambda^5 & \lambda^4 & \lambda^2 \\
 \lambda^3 & \lambda^2 & 1 \\
\end{array}
\right)\ ,
\label{9}
\end{eqnarray}
where we omit the parameters of order unity.
This Yukawa matrix gives the mass ratio for
the up-type quarks as
\begin{eqnarray}
 m_u : m_c : m_t \sim \lambda^6 : \lambda^4 : 1 \ ,
\end{eqnarray}
which is consistent with eq.(\ref{1}) except for 
the up quark but it is maybe within the uncertainty of
parameters.
The prediction for the magnitude of the top quark
Yukawa coupling is also consistent with the 
experimental value $Y_t \sim 1$.

The Yukawa matrix for the
down-type quarks and charged leptons are given by
\begin{eqnarray}
 Y_d = Y_e^T &\sim&
\left(
\begin{array}{ccc} 
\langle \xi_3 \rangle & \langle \xi_2 \rangle  & \langle \xi_1 \rangle 
\end{array} 
\right) 
\left( 
\begin{array}{ccc}  
\lambda^5 & \lambda^4 & \lambda^4 \\
\lambda^4 & \lambda^3 & \lambda^3 \\
\lambda^2 & \lambda & \lambda 
\end{array}
\right)
\left( 
\begin{array}{c}
\langle \xi_3 \rangle^T \\ 
\langle \xi_2 \rangle^T \\
\langle \xi_1 \rangle^T 
\end{array}
\right)/M_*^2 \nonumber \\
&\sim&
\left(
\begin{array}{ccc}  
\lambda^5 & \lambda^4 & \lambda^4 \\
\lambda^4 & \lambda^3 & \lambda^3 \\
\lambda^2 & \lambda & \lambda 
\end{array}
\right)\ .
\label{neo11}
\end{eqnarray}
This Yukawa structure gives
the down-type quark and charged lepton mass ratios as follows:
\begin{eqnarray}
 m_d : m_s : m_b = m_e : m_\mu : m_\tau \sim
 \lambda^4 : \lambda^2 : 1 \ .
\end{eqnarray}
This is consistent with eq.(\ref{1}).
The difference between 
the charged lepton masses and 
down-type quark masses may be explained by the 
introduction of a new Higgs field such as
the SU(5) ${\bf 45}$ representation in the usual way \cite{Harvey}.

Inputting the bottom quark mass,
this model predicts the VEV ratio of the two Higgs doublets as
\begin{eqnarray}
 \tan \beta \equiv 
  \frac{\langle H \rangle}{\langle \bar{H} \rangle} \sim 17\ .
\end{eqnarray}
We obtain the correct
CKM matrix which comes from
Yukawa matrices
(\ref{9}) and (\ref{neo11})
as
\begin{eqnarray}
 V_{\rm CKM} \sim
\left(
\begin{array}{ccc}
 1           & \lambda    & \lambda^3 \\
 - \lambda   & 1          & \lambda^2\\
 - \lambda^3 & -\lambda^2 & 1 \\
\end{array}
\right) \ .
\end{eqnarray}

The interesting point is that
we can obtain the bi-maximal mixing of leptons
without
any fine-tuning in the following mass matrix.
The neutrino mass matrix is proportional to
\begin{equation}
\left( 
\begin{array}{ccc} 
\langle \xi_3 \rangle & \langle \xi_2 \rangle  & \langle \xi_1 \rangle 
\end{array}
\right) 
\left( 
\begin{array}{ccc}  
\lambda^2 & \lambda   & \lambda^3 \\
\lambda   &   1       & \lambda^2 \\
\lambda^3 & \lambda^2 & \lambda^4 
\end{array}
\right)
\left( 
\begin{array}{c}
\langle \xi_3 \rangle^T \\ 
\langle \xi_2 \rangle^T \\
\langle \xi_1 \rangle^T 
\end{array}
\right) 
\sim
\left(
\begin{array}{ccc}
 \lambda^2 & \lambda & \lambda \\
 \lambda   & 1       & 1 \\
 \lambda   & 1       & 1 \\
\end{array}
\right)\ .
\label{15}
\end{equation}
We can see in eq.(\ref{15}) that
the ${\rm SU(3)_H}$ breaking VEVs give
the large mixing for the 2$^{\rm nd}$ and 3$^{\rm rd}$ generations.
Moreover, the 1$^{\rm st}$ and 2$^{\rm nd}$ generation 
mixing is also large.
The easiest way to understand
the large 1$^{\rm st}$ and 2$^{\rm nd}$ generation mixing is 
by changing the basis in eq.(\ref{12}) to
\begin{eqnarray}
 \langle \xi_1 \rangle \sim M_*
\left(
\begin{array}{c}
 0\\
 1\\
 1\\
\end{array}
\right)\ ,\ 
 \langle \xi_2 \rangle \sim M_*
\left(
\begin{array}{c}
 0\\
 0\\
 1\\
\end{array}
\right)\ ,\ 
 \langle \xi_3 \rangle \sim M_*
\left(
\begin{array}{c}
 1\\
 1\\
 1\\
\end{array}
\right)\ .
\label{neo16}
\end{eqnarray}
This change of basis corresponds to
removing the large mixing of 2$^{\rm nd}$ and 3$^{\rm rd}$
generations from 
the mass matrix (\ref{15}).
In this basis, the neutrino mass matrix (\ref{15}) 
is replaced by
\begin{eqnarray}
\left(
\begin{array}{ccc}
 \lambda^2 & \lambda^2 & \lambda \\
 \lambda^2 & \lambda^2 & \lambda \\
 \lambda   & \lambda   & 1 \\
\end{array}
\right)\ .
\label{neo17}
\end{eqnarray}
It turns out that
this matrix gives the large 
1$^{\rm st}$ and 2$^{\rm nd}$ generation mixing.

On the other hand, in this basis,
the charged lepton Yukawa matrix is 
obtained from eq.(\ref{neo11}) 
and eq.(\ref{neo16}) as
\begin{eqnarray}
 Y_e^{\prime} \sim
\left(
\begin{array}{ccc}
 \lambda^5 & \lambda^2 & \lambda^2 \\
 \lambda^4 & \lambda   & \lambda \\
 \lambda^4 & \lambda   & \lambda \\
\end{array}
\right)\ ,
\end{eqnarray}
which gives the large mixing between 
the second and third generations.
Therefore, the lepton mixing matrix is bi-maximal.
The mass ratios of neutrinos are also
suitable for the large MSW solution and are as follows:
\begin{eqnarray}
 m_{\nu 1} : m_{\nu 2} : m_{\nu 3} \sim
\lambda^4 : \lambda^2 : 1 \ .
\end{eqnarray}
The model predicts that
the mixing between the first and the third generations
is of order $\lambda$ and thus small.
It is interesting that
the mixing angle is comparable to the CHOOZ bound,
and will be observed in future long baseline experiments.

This natural derivation of the
bi-maximal mixing is due to the SU(3)$_{\rm H}$ symmetry.
The conventional way 
to obtain the bi-maximal mixing by
an Abelian symmetry 
requires an accidental cancellation in the determinant.
Consider the case that
the neutrino mass matrix is given by
\begin{eqnarray}
\left(
\begin{array}{ccc}
 \lambda^4 & \lambda^2 & \lambda^2 \\
 \lambda^2 & 1         & 1 \\
 \lambda^2 & 1         & 1 \\
\end{array}
\right) \ .
\end{eqnarray}
This type of mass matrix can be given by the 
Abelian flavor symmetry models in which
the second and the third generations
have the same charges.
If there is an accidental cancellation in the
determinant of the 2-3 submatrix so that
the eigenvalues of this submatrix are of order 
$\lambda^2$ and unity,
the matrix in which the 2-3 submatrix is diagonalized
by the large mixing is given by
\begin{eqnarray}
\left(
\begin{array}{ccc}
 \lambda^4 & \lambda^2 & \lambda^2 \\
 \lambda^2 & \lambda^2 & 0  \\
 \lambda^2 & 0         & 1  \\
\end{array}
\right)\ .
\end{eqnarray}
This gives the large $1^{\rm st}$ and $2^{\rm nd}$ generation mixing 
if there is no cancellation in the (1,2) component
while diagonalizing the 2-3 submatrix, and
the correct mass ratio for the MSW solution is reproduced.
However,
without the accidental cancellation,
the mixing angle between the $1^{\rm st}$ and $2^{\rm nd}$ 
generation
is naturally of order $\lambda^2$
and the mass ratios are unacceptable, scaling as $\lambda^2 : 1 : 1$.
As we have seen above,
this cancellation can be 
controlled by the 
non-Abelian horizontal symmetry.

\section{The Model}
\baselineskip 20pt

In this section,
we present a SUSY SU(5) GUT model,
in which the Yukawa structure in the 
previous section is reproduced.

The particle content of this model 
is listed in Tab.\ref{tab1}.
\begin{table}[t]
\begin{center}
\begin{tabular}{|c||c|c|c|c|}\hline
 & ${\rm SU(5)_{GUT}}$ & ${\rm SU(3)_H}$ & ${\rm U(1)_H}$ 
 & ${\rm U(1)_{PQ}}$   \\ \hline\hline
 10              & {\bf 10}       & {\bf 3}         
 & --1/4         & 1 {\rule[-1mm]{0mm}{5mm}\ }    \\ \hline
 $\bar{5}$       & $\bar{\bf 5}$  & {\bf 3}         
 & 3/4           & 1 {\rule[-1mm]{0mm}{5mm}\ }    \\ \hline
 $H$             & {\bf 5}        & {\bf 1}         
 & --1            & 0 {\rule[-1mm]{0mm}{5mm}\ }   \\ \hline
 $\bar{H}$       & $\bar{\bf 5}$  & {\bf 1}         
 & 1             & 0  {\rule[-1mm]{0mm}{5mm}\ }   \\ \hline
 $\xi_1$         & {\bf 1}        & $\bar{\bf 3}$   
 & --1/4          & --1  {\rule[-1mm]{0mm}{5mm}\ }  \\ \hline
 $\xi_2$         & {\bf 1}        & $\bar{\bf 3}$   
 & --9/4          & --1  {\rule[-1mm]{0mm}{5mm}\ }    \\ \hline
 $\xi_3$         & {\bf 1}        & $\bar{\bf 3}$   
 & --13/4         & --1  {\rule[-1mm]{0mm}{5mm}\ }  \\ \hline
 $\bar{\xi_1}$   & {\bf 1}        & ${\bf 3}$       
 & 1/4           & 1  {\rule[-1mm]{0mm}{5mm}\ }    \\ \hline
 $\bar{\xi_2}$   & {\bf 1}        & ${\bf 3}$       
 & 9/4           & 1  {\rule[-1mm]{0mm}{5mm}\ }   \\ \hline
 $\bar{\xi_3}$   & {\bf 1}        & ${\bf 3}$       
 & 13/4          & 1 {\rule[-1mm]{0mm}{5mm}\ }   \\ \hline
 $T$             & {\bf 10}       & {\bf 1}         
 & --1/2          & 0 {\rule[-1mm]{0mm}{5mm}\ }   \\ \hline
 $\bar{T}$       & $\bar{\bf 10}$ & {\bf 1}         
 & 1/2           & 0 {\rule[-1mm]{0mm}{5mm}\ }   \\ \hline
 $F$             & {\bf 5}        & {\bf 1}         
 & 3/2           & 0 {\rule[-1mm]{0mm}{5mm}\ }   \\ \hline
 $\bar{F}$       & $\bar{\bf 5}$  & {\bf 1}         
 & --3/2          & 0  {\rule[-1mm]{0mm}{5mm}\ }   \\ \hline
 $G$             & {\bf 5}        & {\bf 1}         
 & 1/2           & 0 {\rule[-1mm]{0mm}{5mm}\ }   \\ \hline
 $\bar{G}$       & $\bar{\bf 5}$  & {\bf 1}         
 & --1/2          & 0 {\rule[-1mm]{0mm}{5mm}\ }   \\ \hline
 $N$             & {\bf 1}        & {\bf 1}         
 & 5/2           & 0 {\rule[-1mm]{0mm}{5mm}\ }  \\ \hline
 $\Phi$          & {\bf 1}        & {\bf 1}         
 & 1             & 0 {\rule[-1mm]{0mm}{5mm}\ }  \\ \hline
 $\bar{\Phi}$    & {\bf 1}        & {\bf 1}         
 & --1            & 0 {\rule[-1mm]{0mm}{5mm}\ }  \\ \hline
\end{tabular} 
\caption{The particle contents of the SU(5) model.}
\label{tab1} 
\end{center}
\end{table}
The ${\rm U(1)_H}$ charge assignment is an example 
of realistic models.
The ${\rm U(1)_H}$ symmetry
is anomaly free with respect to the ${\rm SU(5)_{GUT}}$,
therefore,
the Nambu-Goldstone boson associated
with ${\rm U(1)_H}$ break down
is purely massless and harmless \cite{Wilczek:1982rv}.

The fields $10$ and $\bar{5}$ are the usual matter fields
as follows:
\begin{eqnarray}
 10 : (q,u^c,e^c) \ \ ,\ \ 
 \bar{5} : (d^c,l) \ .
\end{eqnarray}
The fields $H$ and $\bar{H}$ are the Higgs fields.
The ${\rm SU(2)_L}$ doublet parts of these fields 
remain as
the usual Higgs doublets at the electroweak scale
and their VEVs give the 
masses of the quarks and leptons.
The vector-like fields $T$, $F$, $G$ and $N$ are
the Froggatt-Nielsen (FN) fields which generate the
usual Yukawa interaction terms for quarks and leptons
by being integrated out \cite{Froggatt:1979nt}.
The SU(5) and ${\rm SU(3)_H}$ singlet fields 
$\Phi$ and $\bar{\Phi}$ are
the origin of the mass hierarchy and mixing angle of the 
quarks and leptons 
through the non-renormalizable interactions.

The superpotential which 
creates the Yukawa couplings for the matter fields is
constructed as follows.
\begin{eqnarray}
 W = W_{\rm matter} + W_{\rm Higgs} + W_{\rm mass}\ .
\label{23}
\end{eqnarray}
The matter part $W_{\rm matter}$
contains
${\rm SU(3)_H}$ non-singlet fields.
It is given by
\begin{eqnarray}
 W_{\rm matter} &=& 
\left \{ 
\xi_1   
+ \left( \frac{\Phi}{M_{\rm Pl}} \right)^2 \xi_2
+ \left( \frac{\Phi}{M_{\rm Pl}} \right)^3 \xi_3
\right \}
\cdot 10 \ \bar{T} \nonumber \\
&&
+
\left \{
\left( \frac{\bar{\Phi}}{M_{\rm Pl}} \right)^2 \xi_1
+ \xi_2
+ \left( \frac{\Phi}{M_{\rm Pl}} \right) \xi_3
\right \}
\cdot \bar{5} \ F
\nonumber \\
&&
+
\left \{
\left( \frac{\bar{\Phi}}{M_{\rm Pl}} \right) \xi_1
+ \left( \frac{\Phi}{M_{\rm Pl}} \right) \xi_2
+ \left( \frac{\Phi}{M_{\rm Pl}} \right)^2 \xi_3
\right \}
\cdot \bar{5} \ G
\ ,
\end{eqnarray}
where ` $\cdot$ ' is the inner product of the ${\rm SU(3)_H}$ indices.
We consider the case in which
the fields $\Phi$ and $\bar{\Phi}$ acquire VEVs as follows:
\begin{eqnarray}
 \langle \Phi \rangle = \langle \bar{\Phi} \rangle 
  \sim \lambda M_{\rm Pl} \ .
\end{eqnarray}
In this case,
the ${\rm U(1)_H}$ symmetry breaks down
at the scale of $\lambda M_{\rm Pl}$ and
the factor $\Phi/M_{\rm Pl}$ in the superpotential
can be replaced by the $\lambda \sim 0.22$ which originates the
fermion masses and mixings.

The Higgs part, $W_{\rm Higgs}$, contains the
Yukawa interaction terms of the FN fields and
Higgs fields $H$ and $\bar{H}$ and may be written as
\begin{eqnarray}
 W_{\rm Higgs} =
  T H T
  + \left( \frac{\Phi}{M_{\rm Pl}} \right)
  T \bar{H} \bar{F}
  + T \bar{H} \bar{G}
  + \bar{F} H N
  + \left( \frac{\bar{\Phi}}{M_{\rm Pl}} \right)
  \bar{G} H N \ .
\label{10}
\end{eqnarray}

The last term $W_{\rm mass}$ in eq.(\ref{23})
is the mass term for the FN fields and Higgs fields:
\begin{eqnarray}
 W_{\rm mass} =
 M_* T \bar{T}
  + M_* F \bar{F}
  + M_* G \bar{G}
  + \frac{\bar{\Phi}^5}{M_{\rm Pl}^4}  N N
  + M_* \sum_{i=1}^3 \xi_i \bar{\xi_i}
  + M_{\rm GUT} H \bar{H} \ ,
\label{11}
\end{eqnarray}
where $M_*$ is the FN scale
which can be naturally of the order of $M_{\rm GUT}$
and it can arise from the VEV of a singlet field,
but we do not specify the scale and its origin.
The Majorana mass for the FN field $N$
is given by $\lambda^5 M_{\rm Pl} \sim 10^{14}$ GeV.
It is suitable to give the neutrino masses by
the seesaw mechanism.
The Majorana mass
can be given in another way.
If $M_*$ is obtained by the singlet VEV and
the Planck suppressed mass term in eq.(\ref{11})
is forbidden by some symmetry like $Z_3$,
the Majorana mass term can be given by 
$(\Phi/M_{\rm Pl})^X M_* N N$
and the magnitude can be controlled by
$M_*$ and $X$, which is determined by the ${\rm U(1)_H}$ charge of $N$.
Therefore, hereafter,
we replace the Majorana mass term
with $M_N NN$, where $M_N \sim 10^{14}$ GeV.

We can see that ${\rm U(1)_{PQ}}$ symmetry exists
in this superpotential,
and the charge assignments are listed in Tab.\ref{tab1}.
The non-zero value of the Peccei-Quinn charge of the $\xi$'s
indicates that
the ${\rm SU(3)_H}$ breaking simultaneously induces
the ${\rm U(1)_{PQ}}$ breaking,
which solves the strong CP problem and 
creates the axion dark matter
if the breaking scale $M_*$ 
is around $10^{12}$ GeV \cite{Peccei:1977hh,Kim:1979if,Zhitnitsky:1980tq}.
In this sense,
$M_* \sim M_{\rm GUT}$ may be unacceptable  
due to the creation of too many axions.
A way out is to dilute out them by the inflation of the
universe which is also needed 
by the GUT monopole dilution \cite{Guth:1981zm}.

Now we show that
the fermion masses and mixings are
reproduced by the spontaneous breaking of
the horizontal symmetry.

\begin{figure}
\psfrag{10}{$10$}
\psfrag{xi}{$\xi_1 + \lambda^2 \xi_2 + \lambda^3 \xi_3$}
\psfrag{T}{$T$}
\psfrag{Tb}{$\bar{T}$}
\psfrag{H}{$H$}
\begin{center}
\includegraphics[width=10cm]{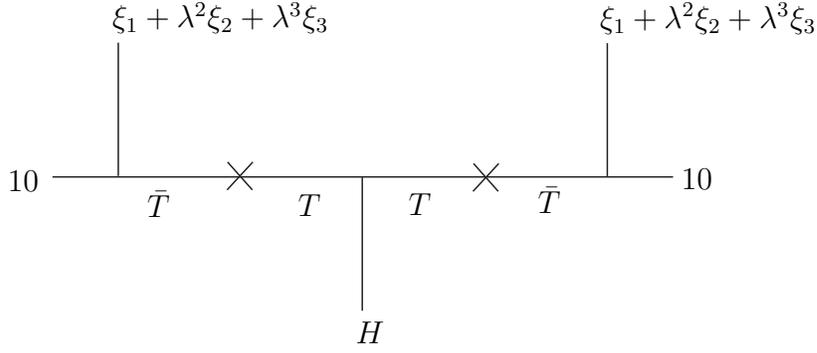} 
\end{center}
\caption{The Feynman diagram for the up-type Yukawa interactions.}
\label{fig1}
\end{figure}

First, let us consider the 
up-type quarks.
The up-type Yukawa couplings are 
given by the Feynman diagram in Fig.\ref{fig1}.
We can extract the Yukawa matrix from this diagram
as follows:
\begin{eqnarray}
 Y_u &\sim& \frac{1}{M_*^2}
\left(
 \lambda^3 \langle \xi_3 \rangle +
 \lambda^2 \langle \xi_2 \rangle +
 \langle \xi_1 \rangle 
\right)
y_T
\left(
 \lambda^3 \langle \xi_3 \rangle +
 \lambda^2 \langle \xi_2 \rangle +
 \langle \xi_1 \rangle 
\right)^T
\nonumber \\
&\sim&
\frac{1}{M_*^2}
\left(
\begin{array}{ccc}
 \langle \xi_3 \rangle & \langle \xi_2 \rangle & \langle \xi_1 \rangle\\
\end{array}
\right)
\left(
\begin{array}{c}
 \lambda^3 \\
 \lambda^2 \\
 1 \\
\end{array}
\right)
y_T
\left(
\begin{array}{ccc}
 \lambda^3 & \lambda^2 & 1 \\
\end{array}
\right)
\left(
\begin{array}{c}
 \langle \xi_3 \rangle^T \\
 \langle \xi_2 \rangle^T \\
 \langle \xi_1 \rangle^T \\
\end{array}
\right)
\ ,
\label{13}
\end{eqnarray}
where $y_T$ is the Yukawa coupling constant
of $THT$ in eq.(\ref{10}).
As we can see in eq.(\ref{13}),
the rank of the $Y_u$ matrix is one, which means 
that only one of the quarks can acquire the 
non-vanishing mass
while the others remain massless.
This situation
usually occurs in the FN mechanism.
In order to avoid this,
more than three pairs of
FN fields $T$ and $\bar{T}$ are necessary.
Then, the coupling $y_T$ is a matrix
and all the components are naturally of order unity 
in the basis that the mass term $M_* T \bar{T}$
is diagonal.
In the case that there are three pairs of $T$ and $\bar{T}$\footnote{
We can obtain the same result for the case where
there are more than three pairs of FN fields.
},
the Yukawa couplings are given by
\begin{eqnarray}
Y_u \!\!\!&\sim&\!\!\!
\frac{1}{M_*^2}
\left(
\begin{array}{ccc}
 \langle \xi_3 \rangle & \langle \xi_2 \rangle & \langle \xi_1 \rangle\\
\end{array}
\right)
\left(
\begin{array}{ccc}
 \lambda^3 & \lambda^3 & \lambda^3 \\
 \lambda^2 & \lambda^2 & \lambda^2 \\
 1         & 1         & 1 \\
\end{array}
\right)
\left(
\begin{array}{ccc}
 1 & 1 & 1 \\
 1 & 1 & 1 \\
 1 & 1 & 1 \\
\end{array}
\right)
\left(
\begin{array}{ccc}
 \lambda^3 & \lambda^2 & 1 \\
 \lambda^3 & \lambda^2 & 1 \\
 \lambda^3 & \lambda^2 & 1 \\
\end{array}
\right)
\left(
\begin{array}{c}
 \langle \xi_3 \rangle^T \\
 \langle \xi_2 \rangle^T \\
 \langle \xi_1 \rangle^T \\
\end{array}
\right)
\nonumber \\
\!\!\!&\sim&\!\!\!
\frac{1}{M_*^2}
\left(
\begin{array}{ccc}
 \langle \xi_3 \rangle & \langle \xi_2 \rangle & \langle \xi_1 \rangle\\
\end{array}
\right)
\left(
\begin{array}{ccc}
 \lambda^6 & \lambda^5 & \lambda^3 \\
 \lambda^5 & \lambda^4 & \lambda^2 \\
 \lambda^3 & \lambda^2 & 1         \\
\end{array}
\right)
\left(
\begin{array}{c}
 \langle \xi_3 \rangle^T \\
 \langle \xi_2 \rangle^T \\
 \langle \xi_1 \rangle^T \\
\end{array}
\right)
\ .
\label{14}
\end{eqnarray}
The second and the fourth matrix in the first expression in eq.(\ref{14})
is the coupling matrix between $10$ and $\bar{T}$ fields
and the center matrix is the $y_T$ matrix.
This $Y_u$ matrix has the same structure as eq.(\ref{7}).

\begin{figure}
\psfrag{10}{$10$}
\psfrag{xi}{$\xi_1 + \lambda^2 \xi_2 + \lambda^3 \xi_3$}
\psfrag{xi2g}{$\lambda \xi_1 + \lambda \xi_2 + \lambda^2 \xi_3$}
\psfrag{xi2f}{$\lambda^2 \xi_1 + \xi_2 + \lambda \xi_3$}
\psfrag{T}{$T$}
\psfrag{Tb}{$\bar{T}$}
\psfrag{Hb}{${\bar{H}}$}
\psfrag{G}{$G$}
\psfrag{Gb}{$\bar{G}$}
\psfrag{F}{$F$}
\psfrag{Fb}{$\bar{F}$}
\psfrag{lam}{$\lambda$}
\psfrag{5b}{$\bar{5}$}
\begin{center}
\includegraphics[width=10cm]{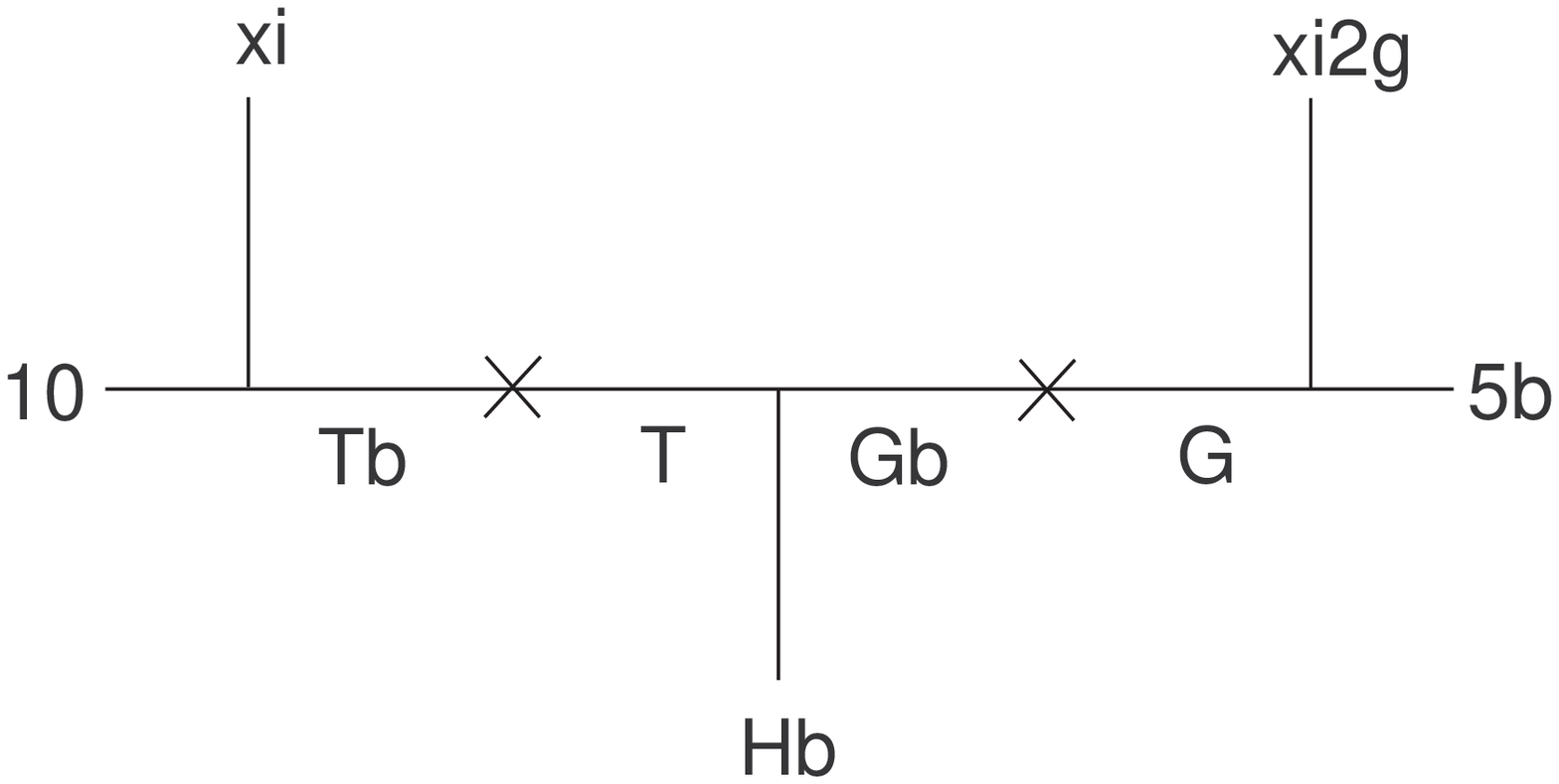} 
\includegraphics[width=10cm]{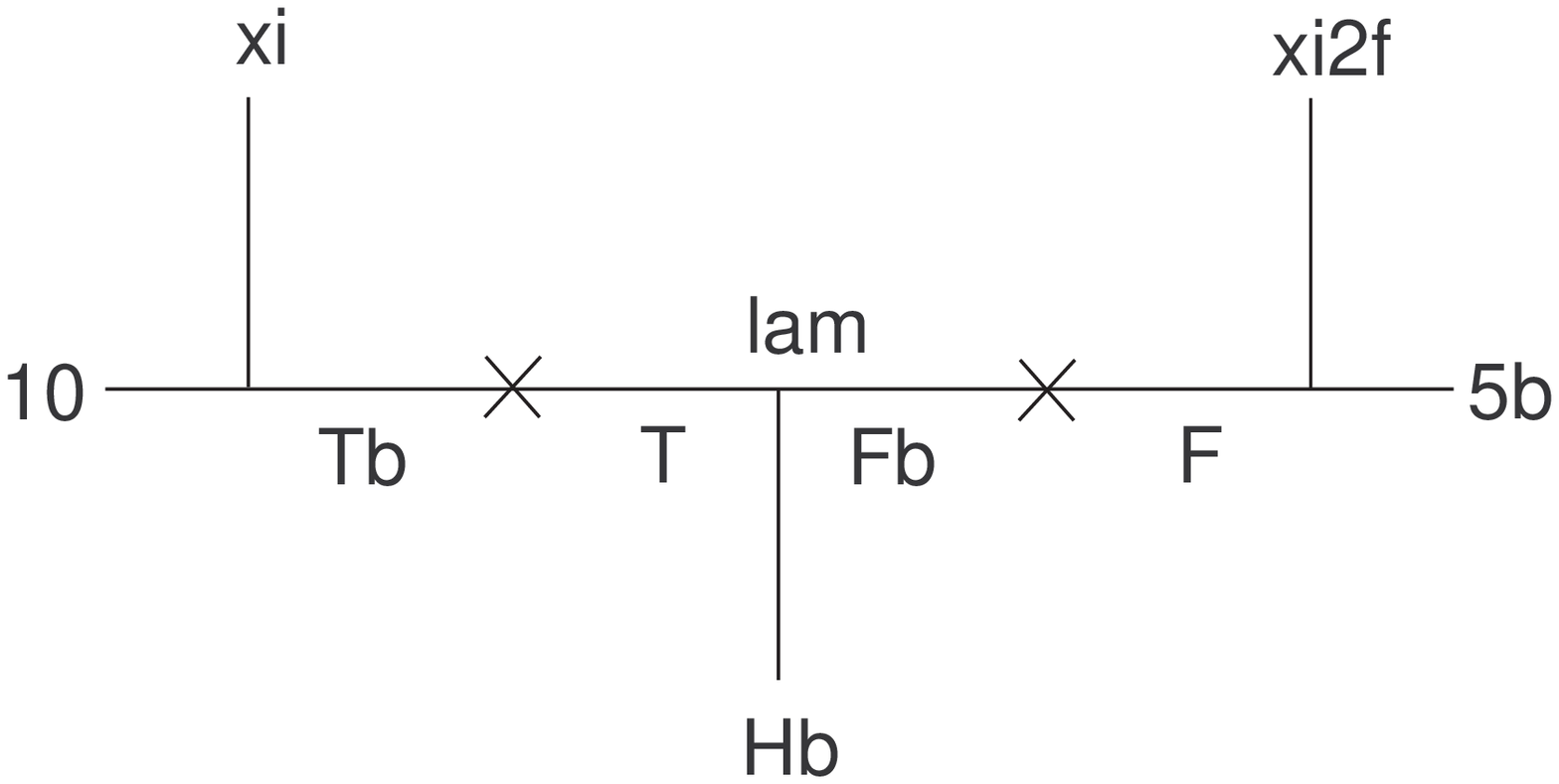}
\end{center}
\caption{The Feynman diagrams for the down-type and
charged lepton Yukawa interactions.}
\label{fig2}
\end{figure}

For the down-type quarks and charged leptons,
there are two Feynman diagrams because 
we introduce
two kinds of ${\rm SU(5)_{GUT}}$
${\bf 5}$ representation FN fields $F$ and $G$,
and again more than three pairs of these fields are
necessary.
The introduction of two kinds of ${\bf 5}$ fields
is the essential point of this model.
The field $G$ gives the down-type quarks and charged leptons
mass matrices and $F$ gives the neutrino mass matrix
suitably without disturbing each other.
The two diagrams are shown in Fig.\ref{fig2}.
From the upper diagram, which is $G$ contribution,
the Yukawa structure is given as
\begin{eqnarray}
Y_d^G \!\!\!&=&\!\!\! (Y_e^{G})^T 
\nonumber \\
\!\!\!&\sim&\!\!\!
\frac{1}{M_*^2}
\left(
\begin{array}{ccc}
 \langle \xi_3 \rangle & \langle \xi_2 \rangle & \langle \xi_1 \rangle\\
\end{array}
\right)
\left(
\begin{array}{ccc}
 \lambda^3 & \lambda^3 & \lambda^3 \\
 \lambda^2 & \lambda^2 & \lambda^2 \\
 1         & 1         & 1 \\
\end{array}
\right)
\left(
\begin{array}{ccc}
 1 & 1 & 1 \\
 1 & 1 & 1 \\
 1 & 1 & 1 \\
\end{array}
\right)
\left(
\begin{array}{ccc}
 \lambda^2 & \lambda & \lambda \\
 \lambda^2 & \lambda & \lambda \\
 \lambda^2 & \lambda & \lambda \\
\end{array}
\right)
\left(
\begin{array}{c}
 \langle \xi_3 \rangle^T \\
 \langle \xi_2 \rangle^T \\
 \langle \xi_1 \rangle^T \\
\end{array}
\right)
\nonumber \\
\!\!\!&\sim&\!\!\!
\frac{1}{M_*^2}
\left(
\begin{array}{ccc}
 \langle \xi_3 \rangle & \langle \xi_2 \rangle & \langle \xi_1 \rangle\\
\end{array}
\right)
\left(
\begin{array}{ccc}
 \lambda^5 & \lambda^4 & \lambda^4 \\
 \lambda^4 & \lambda^3 & \lambda^3 \\
 \lambda^2 & \lambda   & \lambda   \\
\end{array}
\right)
\left(
\begin{array}{c}
 \langle \xi_3 \rangle^T \\
 \langle \xi_2 \rangle^T \\
 \langle \xi_1 \rangle^T \\
\end{array}
\right)\ .
\label{16}
\end{eqnarray}
The $F$ contribution (the lower diagram)
is
\begin{eqnarray}
Y_d^F \!\!\!&=&\!\!\! (Y_e^{F})^T 
\nonumber \\
\!\!\!&\sim&\!\!\!
\frac{1}{M_*^2}
\left(
\begin{array}{ccc}
 \langle \xi_3 \rangle & \langle \xi_2 \rangle & \langle \xi_1 \rangle\\
\end{array}
\right)
\left(
\begin{array}{ccc}
 \lambda^3 & \lambda^3 & \lambda^3 \\
 \lambda^2 & \lambda^2 & \lambda^2 \\
 1         & 1         & 1 \\
\end{array}
\right)
\left(
\begin{array}{ccc}
 \lambda & \lambda & \lambda \\
 \lambda & \lambda & \lambda \\
 \lambda & \lambda & \lambda \\
\end{array}
\right)
\left(
\begin{array}{ccc}
 \lambda & 1 & \lambda^2 \\
 \lambda & 1 & \lambda^2 \\
 \lambda & 1 & \lambda^2 \\
\end{array}
\right)
\left(
\begin{array}{c}
 \langle \xi_3 \rangle^T \\
 \langle \xi_2 \rangle^T \\
 \langle \xi_1 \rangle^T \\
\end{array}
\right)
\nonumber \\
\!\!\!&\sim&\!\!\!
\frac{1}{M_*^2}
\left(
\begin{array}{ccc}
 \langle \xi_3 \rangle & \langle \xi_2 \rangle & \langle \xi_1 \rangle\\
\end{array}
\right)
\left(
\begin{array}{ccc}
 \lambda^5 & \lambda^4 & \lambda^6 \\
 \lambda^4 & \lambda^3 & \lambda^5 \\
 \lambda^2 & \lambda   & \lambda^3   \\
\end{array}
\right)
\left(
\begin{array}{c}
 \langle \xi_3 \rangle^T \\
 \langle \xi_2 \rangle^T \\
 \langle \xi_1 \rangle^T \\
\end{array}
\right)\ .
\label{17}
\end{eqnarray}
Then,
the Yukawa couplings are given as follows:
\begin{eqnarray}
 Y_d = Y_e^T \sim Y_d^G + Y_d^F \sim 
\frac{1}{M_*^2}
\left(
\begin{array}{ccc}
 \langle \xi_3 \rangle & \langle \xi_2 \rangle & \langle \xi_1 \rangle\\
\end{array}
\right)
\left(
\begin{array}{ccc}
 \lambda^5 & \lambda^4 & \lambda^4 \\
 \lambda^4 & \lambda^3 & \lambda^3 \\
 \lambda^2 & \lambda   & \lambda   \\
\end{array}
\right)
\left(
\begin{array}{c}
 \langle \xi_3 \rangle^T \\
 \langle \xi_2 \rangle^T \\
 \langle \xi_1 \rangle^T \\
\end{array}
\right)
\ .
\label{18}
\end{eqnarray}
This reproduces the Yukawa structure in eq.(\ref{neo11}).

\begin{figure}[htbp]
\psfrag{H}{$H$}
\psfrag{N}{$N$}
\psfrag{xi2g}{$\lambda \xi_1 + \lambda \xi_2 + \lambda^2 \xi_3$}
\psfrag{xi2f}{$\lambda^2 \xi_1 + \xi_2 + \lambda \xi_3$}
\psfrag{G}{$G$}
\psfrag{Gb}{$\bar{G}$}
\psfrag{F}{$F$}
\psfrag{Fb}{$\bar{F}$}
\psfrag{lam}{$\lambda$}
\psfrag{5b}{$\bar{5}$}
\psfrag{MN}{$M_N$}
\begin{center}
\includegraphics[width=10cm]{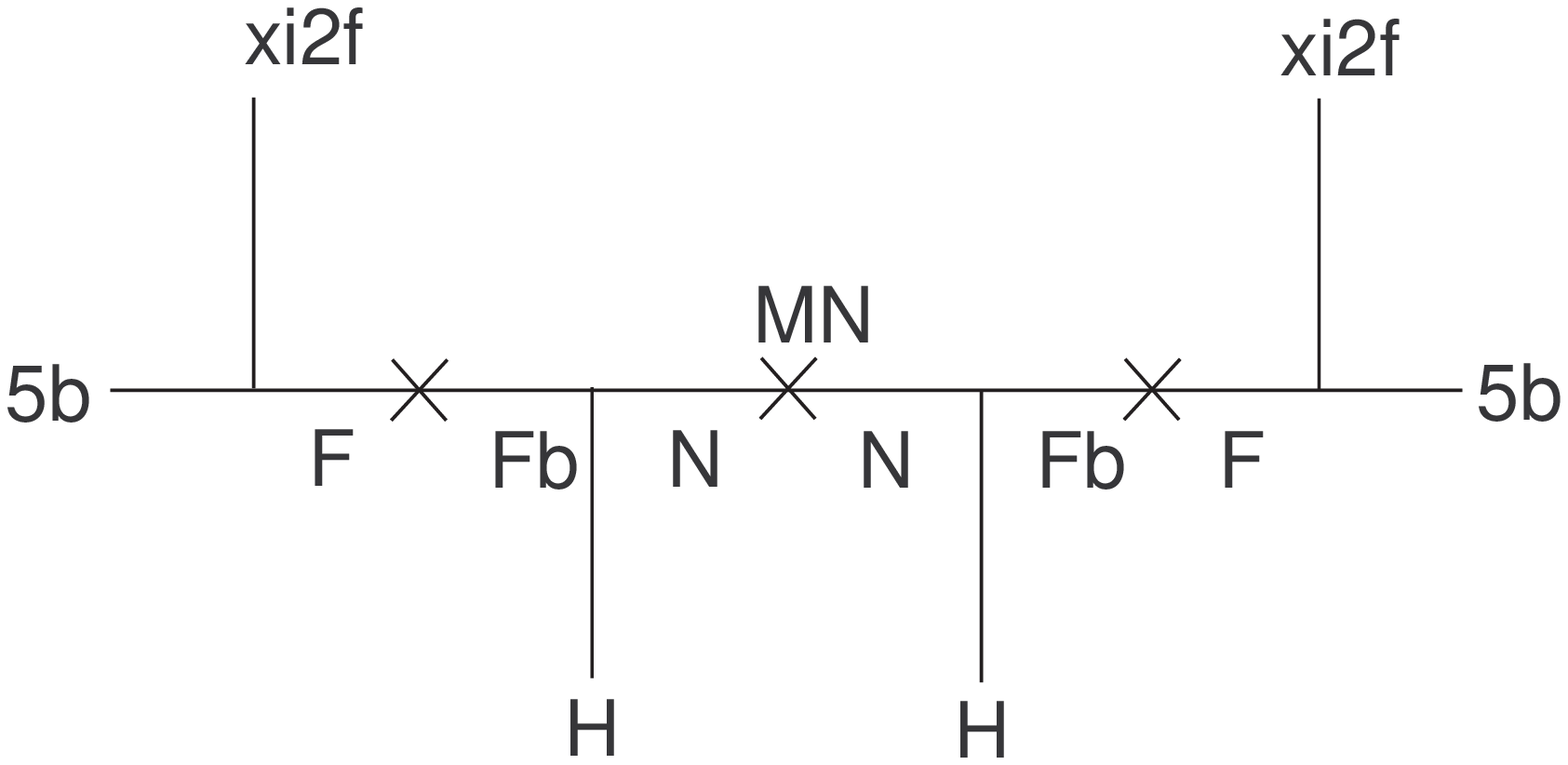} 
\includegraphics[width=10cm]{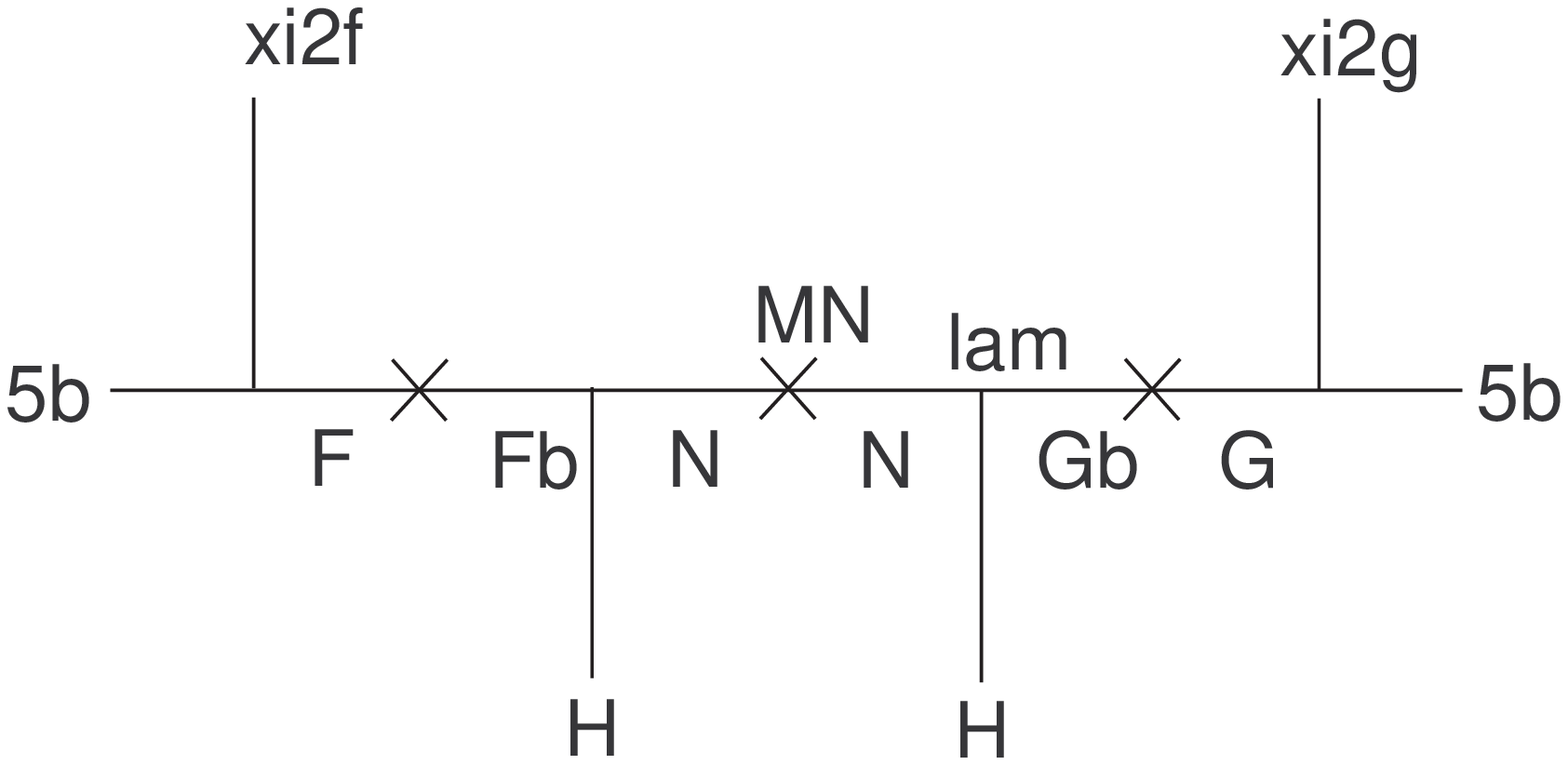} 
\includegraphics[width=10cm]{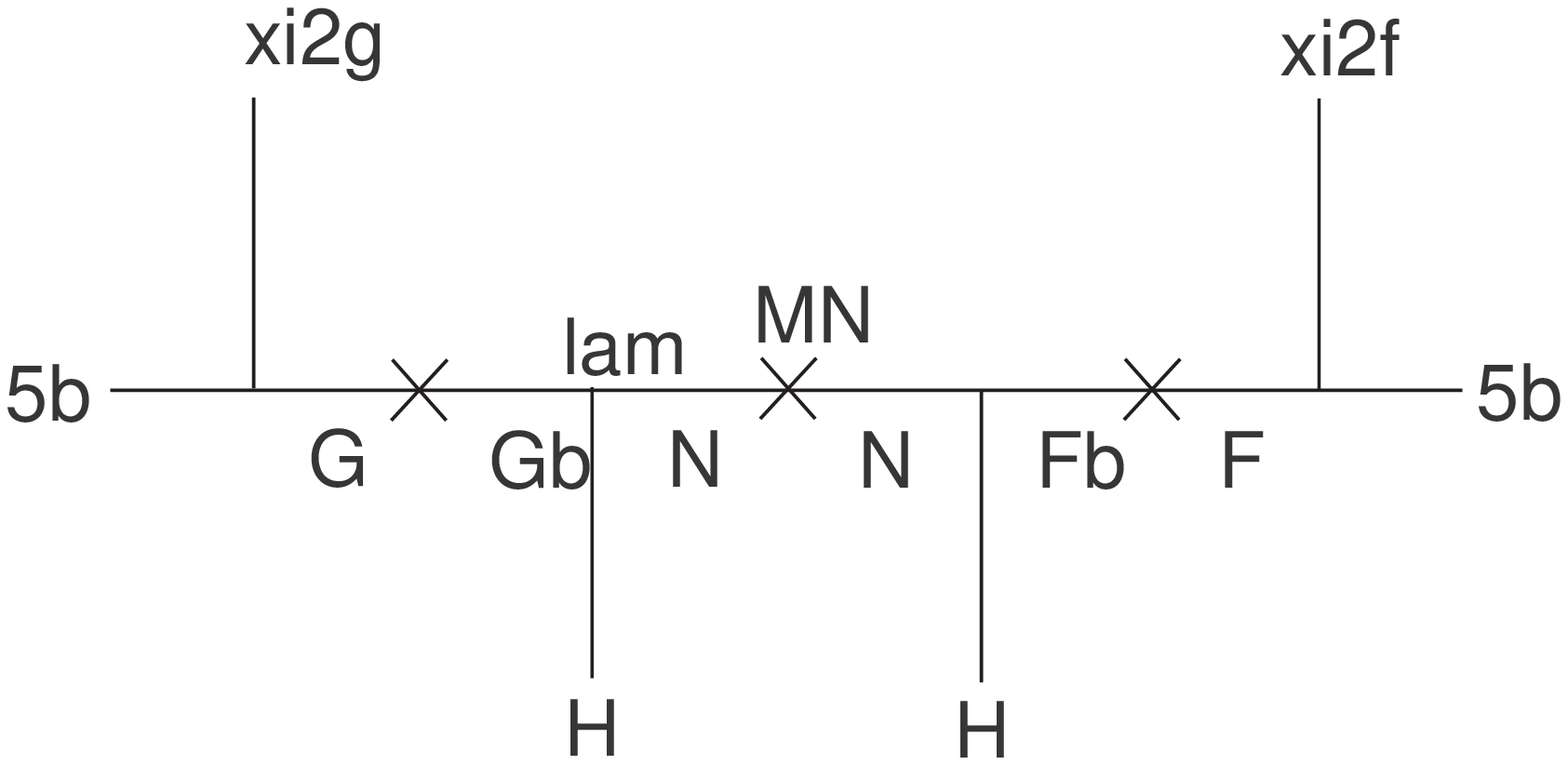} 
\includegraphics[width=10cm]{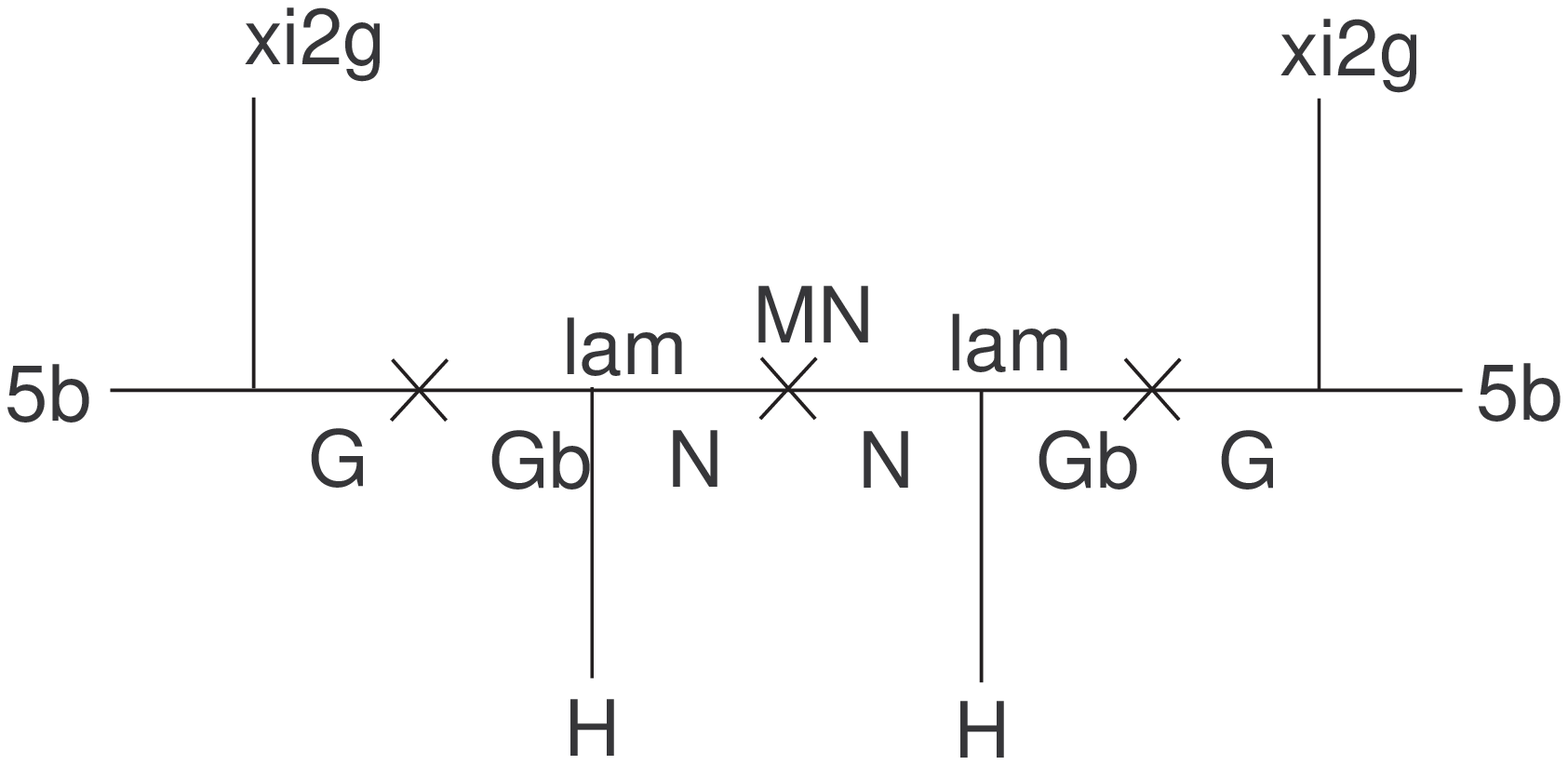} 
\end{center}
\caption{The Feynman diagrams for the neutrino masses.}
\label{fig3}
\end{figure}

Now, let us inspect the 
neutrino mass matrix
whose entries originate from
the Feynman diagrams shown in Fig.\ref{fig3}.
The first diagram gives the main contribution:
\begin{equation}
 m_\nu^{FF} 
 \sim \frac{\langle H \rangle^2}{M_N}
\left(
\begin{array}{ccc}
 \langle \xi_3 \rangle & \langle \xi_2 \rangle & \langle \xi_1 \rangle\\
\end{array}
\right)
\left(
\begin{array}{ccc}
 \lambda^2 & \lambda & \lambda^3 \\
 \lambda   & 1       & \lambda^2 \\
 \lambda^3   & \lambda^2       & \lambda^4 \\
\end{array}
\right)
\left(
\begin{array}{c}
 \langle \xi_3 \rangle^T \\
 \langle \xi_2 \rangle^T \\
 \langle \xi_1 \rangle^T \\
\end{array}
\right)
\ .
\label{24}
\end{equation}
The others are given as follows.
\begin{equation}
 m_\nu^{FG} 
\sim \frac{\langle H \rangle^2}{M_N}
\left(
\begin{array}{ccc}
 \langle \xi_3 \rangle & \langle \xi_2 \rangle & \langle \xi_1 \rangle\\
\end{array}
\right)
\left(
\begin{array}{ccc}
 \lambda^4 & \lambda^3 & \lambda^3 \\
 \lambda^3   & \lambda^2       & \lambda^2 \\
 \lambda^3   & \lambda^2       & \lambda^2 \\
\end{array}
\right)
\left(
\begin{array}{c}
 \langle \xi_3 \rangle^T \\
 \langle \xi_2 \rangle^T \\
 \langle \xi_1 \rangle^T \\
\end{array}
\right)
\ ,
\label{34}
\end{equation}
\begin{equation}
m_\nu^{GF} \sim (m_\nu^{FG})^T \ ,
\label{35}
\end{equation}
\begin{equation}
 m_\nu^{GG} 
\sim \frac{\langle H \rangle^2}{M_N}
\left(
\begin{array}{ccc}
 \langle \xi_3 \rangle & \langle \xi_2 \rangle & \langle \xi_1 \rangle\\
\end{array}
\right)
\left(
\begin{array}{ccc}
 \lambda^6 & \lambda^5 & \lambda^5 \\
 \lambda^5   & \lambda^4       & \lambda^4 \\
 \lambda^5   & \lambda^4       & \lambda^4 \\
\end{array}
\right)
\left(
\begin{array}{c}
 \langle \xi_3 \rangle^T \\
 \langle \xi_2 \rangle^T \\
 \langle \xi_1 \rangle^T \\
\end{array}
\right)
\ .
\label{36}
\end{equation}
The contributions from
eq.(\ref{34}--\ref{36})
are small compared to $m_\nu^{FF}$ and can be safely ignored.
The mass matrix $m_\nu \sim m_\nu^{FF}$ gives the 
bi-maximal neutrino mixing as shown in the previous section.

One may think that
the $G$ contribution is not necessary
if the $F$ contribution gives the neutrino mass matrix suitably.
However the large mixing between the second and third generations
does not occur 
in absence of the $G$ fields.
This issue is easily seen in the basis of eq.(\ref{neo16}).
In this basis, the charged lepton mass matrix
can be read off from eq.(\ref{17}) as follows,
\begin{equation}
Y_e^{\prime F} \sim \left(
            \begin{array}{ccc}
            \lambda^5 & \lambda^2 & \lambda^2 \\
            \lambda^5 & \lambda^2 & \lambda^2 \\
            \lambda^4 & \lambda   & \lambda   
            \end{array}
            \right) \ .
\end{equation}
This charged lepton mass matrix 
and the neutrino mass matrix in eq.(\ref{neo17})
do not contain the large $2^{\rm nd}$ and $3^{\rm rd}$
generation mixing.

\section{SO(10) Embedding}

In this section,
we construct the ${\rm SO(10)_{GUT}}$ model in which
the matter fields are 
unified in a ${\bf 16}$ representation.
This matter unification is 
an attractive feature but
this is the difficulty of the 
${\rm SO(10)_{GUT}}$ model building simultaneously.

The minimal ${\rm SO(10)_{GUT}}$ model is not realistic
because the Yukawa couplings for the up-type and down-type quarks
coincide.
To avoid this,
we extend the Higgs sector to introduce the
{\bf 16} representation Higgs field $H_{16}$ whose 
SU(5) ${\bf \bar{5}}$ component mixes with that of 
usual Higgs field $H_{10}$,
i.e.\
the down type quark and charged lepton masses are
both given by the VEV of $H_{16}$ and $H_{10}$.
This situation can be realized by the introduction of 
$\chi$ and $\bar{\chi}$ fields which are {\bf 16} and ${\bf \bar{16}}$ 
representations of SO(10), respectively, 
and break the SO(10) symmetry to SU(5) by
the SU(5) singlet components acquire the VEV.
This idea has been considered in Ref.\cite{Albright:1998vf, Nomura:1999gm}.
We use this mechanism and
we can reproduce the
appropriate quark masses, quark mixings, 
charged lepton masses, neutrino masses and mixings for
the large angle MSW solution
by imposing the SU(3) $\times$ U(1) horizontal symmetry.

The particle content and 
their horizontal charges are 
listed in Tab.\ref{tab2}.
\begin{table}[t]
\begin{center}
\begin{tabular}{|c||c|c|c|c|}\hline
 & ${\rm SO(10)_{GUT}}$ & ${\rm SU(3)_H}$ & ${\rm U(1)_H}$ 
 & ${\rm U(1)_{PQ}}$   \\ \hline\hline
 16              & {\bf 16}       & {\bf 3}         
 & 0             & 1 {\rule[-1mm]{0mm}{5mm}\ }  \\ \hline
 $H_{10}$        & {\bf 10}       & {\bf 1}         
 & --11/2         & 0 {\rule[-1mm]{0mm}{5mm}\ }  \\ \hline
 $H_{16}$        & {\bf 16}       & {\bf 1}         
 & --13/4         & 0 {\rule[-1mm]{0mm}{5mm}\ }  \\ \hline
 $\bar{H}_{\bar{16}}$  & ${\bf \bar{16}}$   & {\bf 1}         
 & 13/4             & 0 {\rule[-1mm]{0mm}{5mm}\ }  \\ \hline
 $\xi_1$         & {\bf 1}        & $\bar{\bf 3}$   
 & 11/4           & --1 {\rule[-1mm]{0mm}{5mm}\ }   \\ \hline
 $\xi_2$         & {\bf 1}        & $\bar{\bf 3}$   
 & 3/4         & --1 {\rule[-1mm]{0mm}{5mm}\ }  \\ \hline
 $\xi_3$         & {\bf 1}        & $\bar{\bf 3}$   
 & --1/4         & --1 {\rule[-1mm]{0mm}{5mm}\ }   \\ \hline
 $\bar{\xi_1}$   & {\bf 1}        & ${\bf 3}$       
 & --11/4         & 1 {\rule[-1mm]{0mm}{5mm}\ }   \\ \hline
 $\bar{\xi_2}$   & {\bf 1}        & ${\bf 3}$       
 & --3/4           & 1 {\rule[-1mm]{0mm}{5mm}\ }  \\ \hline
 $\bar{\xi_3}$   & {\bf 1}        & ${\bf 3}$       
 & 1/4           & 1 {\rule[-1mm]{0mm}{5mm}\ } \\ \hline
 $S$             & {\bf 16}       & {\bf 1}         
 & 11/4           & 0 {\rule[-1mm]{0mm}{5mm}\ }  \\ \hline
 $\bar{S}$       & ${\bf \bar{16}}$ & {\bf 1}         
 & --11/4           & 0 {\rule[-1mm]{0mm}{5mm}\ } \\ \hline
 $J$             & {\bf 10}       & {\bf 1}         
 &  1/2          & 0 {\rule[-1mm]{0mm}{5mm}\ }   \\ \hline
 $\chi$          & {\bf 16}       & {\bf 1}         
 & --5/4           & 0 {\rule[-1mm]{0mm}{5mm}\ }  \\ \hline
 $\bar{\chi}$   & ${\bf \bar{16}}$ & {\bf 1}         
 & 5/4            & 0 {\rule[-1mm]{0mm}{5mm}\ }  \\ \hline
 $\Phi$          & {\bf 1}        & {\bf 1}         
 & 1             & 0 {\rule[-1mm]{0mm}{5mm}\ }   \\ \hline
 $\bar{\Phi}$    & {\bf 1}        & {\bf 1}         
 & --1            & 0 {\rule[-1mm]{0mm}{5mm}\ } \\ \hline
\end{tabular} 
\caption{The particle contents of the SO(10) model.}
\label{tab2} 
\end{center}
\end{table}
In this model,
the SO(10)$_{\rm GUT}$ and ${\rm U(1)_H}$
is broken spontaneously by
the VEVs of the fields $\chi$, $\bar{\chi}$, $\Phi$ and $\bar{\Phi}$
at the order of $\lambda M_{\rm Pl}$.
Below this scale, 
the ${\rm SU(5)_{GUT}} \times {\rm SU(3)_H}$ symmetry
remains.

First, we consider the 
Higgs sector.
The relevant superpotential for the Higgs fields are given by
\begin{eqnarray}
W =
M_H H_{16} \bar{H}_{\bar{16}}
+ \left( \frac{{\Phi}}{M_{\rm Pl}} \right)
H_{10} \bar{H}_{\bar{16}} \bar{\chi}
\end{eqnarray}
The breaking of ${\rm SO(10)_{GUT}}$ 
and ${\rm U(1)_H}$
leads to the 
term $\lambda \langle \bar{\chi} \rangle
H_{10} \bar{H}_{\bar{16}}$.
If $M_H$ is comparable to 
$\lambda \langle \bar \chi \rangle \sim
\lambda^2 M_{\rm Pl} \sim 10^{17}$ GeV,
the SU(5) ${\bf \bar{5}}$ component of
$H_{10}$ and that of $H_{16}$ are strongly mixed, and
the down-type quark and charged lepton masses
are given by both $H_{10}$ and $H_{16}$.
This assumption of the Higgs mass scale is suitable for 
suppressing proton decay \cite{Lucas:1997bc}, 
although this mass scale is slightly larger than the
ordinary grand unified scale, which is about $10^{16}$ GeV.
We assume such a mass scale $M_H \sim 10^{17}$ GeV.

The matter interactions are given as follows:
\begin{eqnarray}
W_{\rm matter} &=&
\left \{
\xi_1 
+ \left( \frac{\Phi}{M_{\rm Pl}} \right)^2 \xi_2
+ \left( \frac{\Phi}{M_{\rm Pl}} \right)^3 \xi_3
\right \} \cdot 16 \ \bar{S} \nonumber \\
&&
+
\frac{\chi}{M_{\rm Pl}}
\left \{
\left( \frac{\bar{\Phi}}{M_{\rm Pl}} \right)^2 \xi_1
+ \xi_2
+ \left( \frac{\Phi}{M_{\rm Pl}} \right) \xi_3
\right \} \cdot 16 \ J\ .
\label{37}
\end{eqnarray} 
Below the SO(10) breaking scale,
$\chi/M_{\rm Pl}$, $\Phi/M_{\rm Pl}$ and $\bar{\Phi}/M_{\rm Pl}$
can be replaced with $\lambda \sim 0.22$.

The Yukawa interaction terms of FN fields 
are given by
\begin{eqnarray}
 W_{\rm FN} =
SS H_{10} + SJ H_{16} \ .
\end{eqnarray}
The other non-renormalizable interaction terms are small and 
they do not give the leading contribution to fermion masses and mixings.
The mass terms of FN fields are given by
\begin{eqnarray}
W_{\rm mass} &=&
\left( \frac{\bar{\Phi}}{M_{\rm Pl}} \right)^2 SJ \chi
+ \left( \frac{{\Phi}}{M_{\rm Pl}} \right) \bar{S} J \bar{\chi}
+ M_* S \bar{S}
+ \left( \frac{\bar{\Phi}}{M_{\rm Pl}} \right) M_* JJ 
\nonumber \\
&&
+ \left( \frac{\bar{\Phi}}{M_{\rm Pl}} \right)^8 
SS \bar{\chi} \bar{\chi} 
+ \left( \frac{{\Phi}}{M_{\rm Pl}} \right)^8 
\bar{S} \bar{S} {\chi} {\chi} 
\ .
\label{41}
\end{eqnarray}
By the SO(10) breaking,
the first two terms give the mass terms for 
SU(5) {\bf 5} component of $\bar{S}$ ($5_{\bar{S}}$) and $J$ ($5_{J}$)
and SU(5) ${\bf \bar{5}}$ component of $S$ ($\bar{5}_{S}$)
and $J$ ($\bar{5}_J$) as follows:
\begin{eqnarray}
 \lambda^3 M_{\rm Pl} \bar{5}_{S} 5_J 
+ \lambda^2 M_{\rm Pl} 5_{\bar{S}} \bar{5}_J\ .
\label{42}
\end{eqnarray}
If $M_* \sim M_{\rm GUT}$, 
the contribution of the next two terms in eq.(\ref{41}) is negligible
compared to eq.(\ref{42}) and
the mass eigenstates for the {\bf 5} and ${\bf \bar{5}}$
FN fields are ($\bar{S}$, $J$) 
and ($J$, $S$) pair.
It is true for sufficiently small $M_*$ 
($M_* \lesssim M_{\rm GUT}$),
but as we will show later
small $M_*$ leads to small Yukawa couplings
for down-type quarks which needs small $\tan \beta$ 
($M_* \gtrsim 10^{15}$ GeV).
The last two terms of eq.(\ref{41}) give the Majorana mass terms
for SU(5) singlet FN fields.

\begin{figure}
\psfrag{16}{$16$}
\psfrag{xi}{$\xi_1 + \lambda^2 \xi_2 + \lambda^3 \xi_3$}
\psfrag{S}{$S$}
\psfrag{Sb}{$\bar{S}$}
\psfrag{H10}{$H_{10}$}
\begin{center}
\includegraphics[width=10cm]{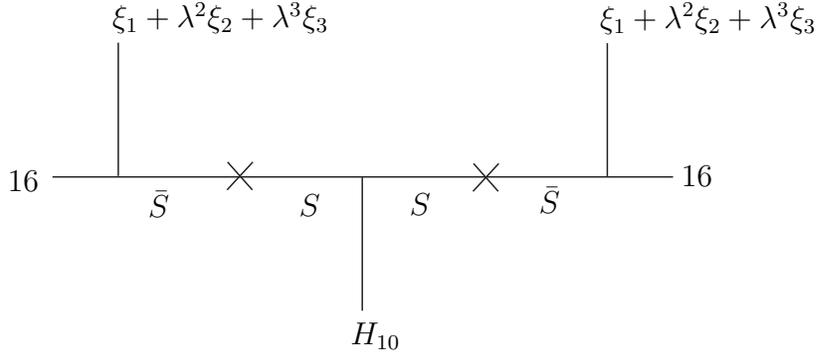} 
\end{center}
\caption{The Feynman diagram for the up-type Yukawa interactions
in the SO(10) model.}
\label{fig4}
\end{figure}

Now we construct the Yukawa structures.
The Yukawa couplings for the up-type quarks are
given by the Feynman diagram in Fig.\ref{fig4}.
Since this diagram is similar to the SU(5) case (Fig.\ref{fig1}),
the desired Yukawa structure is given as in eq.(\ref{14}).

\begin{figure}
\psfrag{16}{$16$}
\psfrag{xi}{$\xi_1 + \lambda^2 \xi_2 + \lambda^3 \xi_3$}
\psfrag{xi2f}{$\lambda^3 \xi_1 + \lambda \xi_2 + \lambda^2 \xi_3$}
\psfrag{S}{$S$}
\psfrag{Sb}{$\bar{S}$}
\psfrag{J}{$J$}
\psfrag{H10}{$H_{10}$}
\psfrag{H16}{$H_{16}$}
\psfrag{lam2}{$\lambda^2 M_{\rm Pl}$}
\psfrag{lam3}{$\lambda^3 M_{\rm Pl}$}
\begin{center}
\includegraphics[width=10cm]{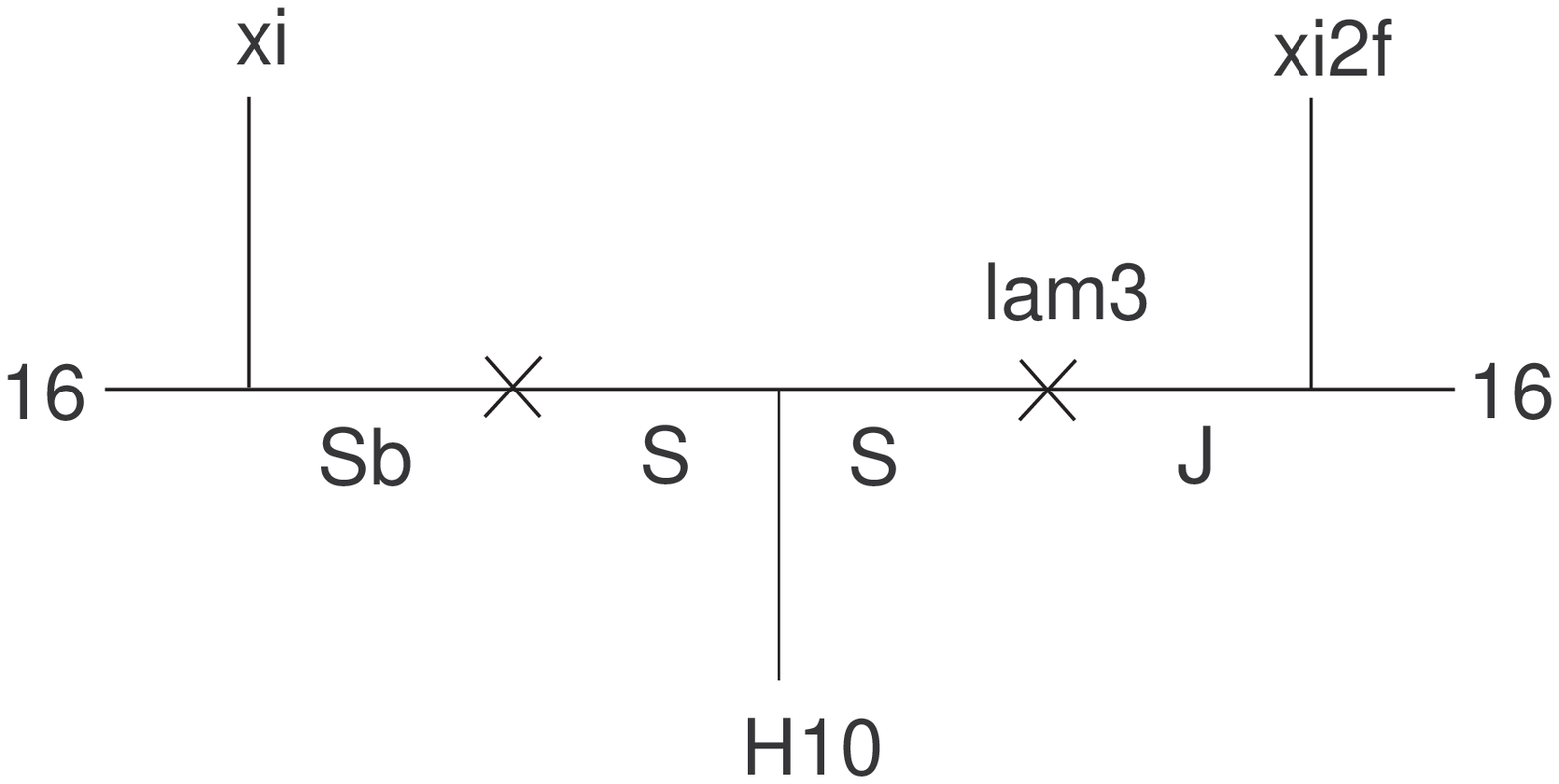} 
\includegraphics[width=10cm]{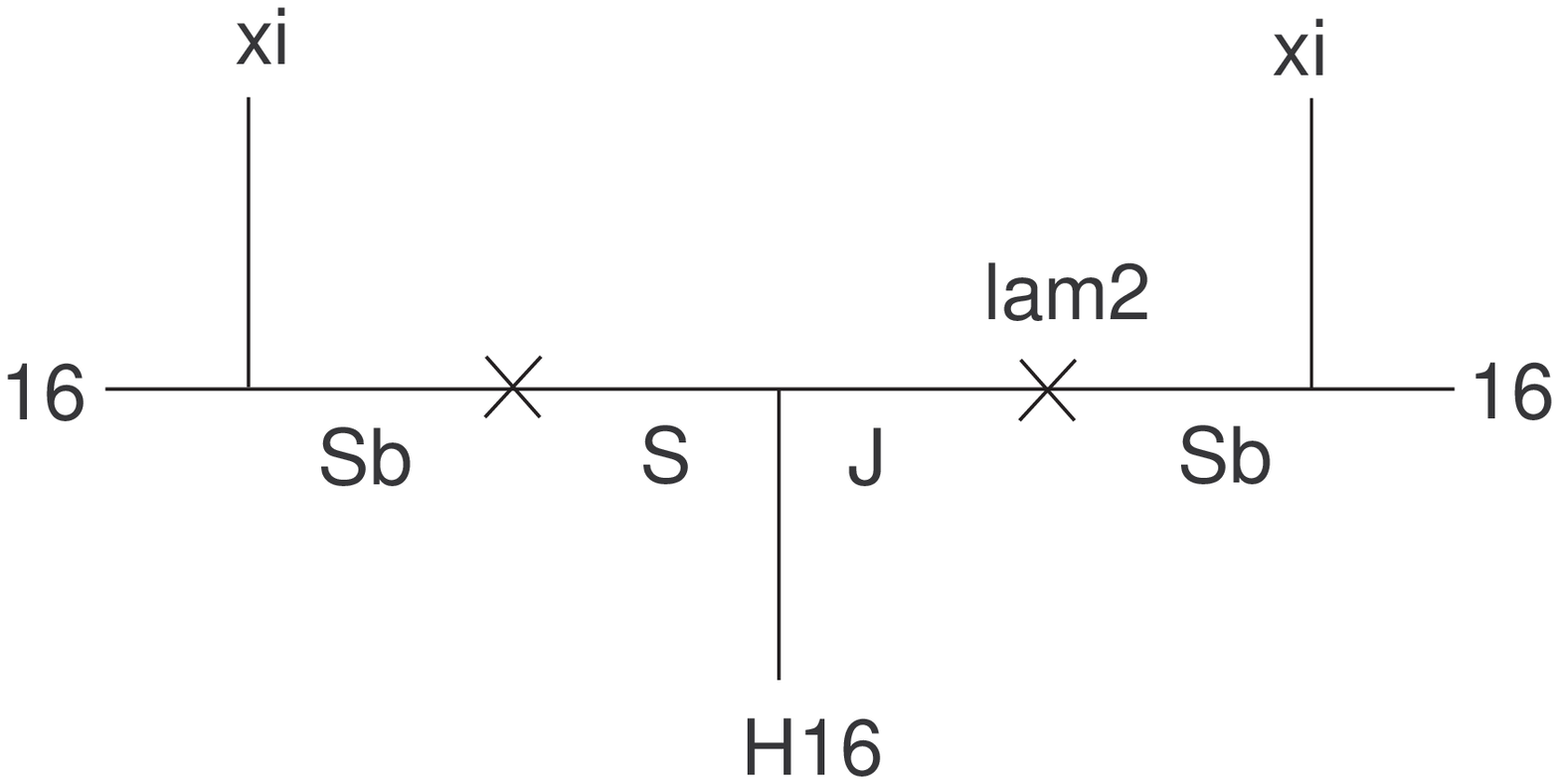} 
\end{center}
\caption{The Feynman diagrams for the down-type quark and
charged lepton Yukawa interactions
in the SO(10) model.}
\label{fig5}
\end{figure}

The down-type quark and the charged lepton Yukawa structures are given by 
the sum of the two contributions in Fig.\ref{fig5}.
The first diagram gives the Yukawa coupling as follows:
\begin{eqnarray}
 Y_d^{(1)} = (Y_e^{(1)})^T \sim
\frac{M_*}{\lambda^3 M_{\rm Pl}}
\left(
\begin{array}{ccc}
 \langle \xi_3 \rangle & \langle \xi_2 \rangle & \langle \xi_1 \rangle \\
\end{array}
\right)
\left(
\begin{array}{ccc}
 \lambda^5 & \lambda^4 & \lambda^6 \\
 \lambda^4 & \lambda^3 & \lambda^5 \\
 \lambda^2 & \lambda   & \lambda^3 \\
\end{array}
\right) 
\left(
\begin{array}{c}
 \langle \xi_3 \rangle^T \\
 \langle \xi_2 \rangle^T \\
 \langle \xi_1 \rangle^T \\
\end{array}
\right)/M_*^2
\ .
\end{eqnarray}
Only with this matrix,
it is unwanted type as in the case of $F$ contribution (eq.(\ref{17}))
in the ${\rm SU(5)_{GUT}}$ model 
because this type does not give 
the maximal mixing for atmospheric neutrinos.
However, by adding another contribution 
we can obtain the maximal mixing.
It is given by the second diagram in Fig.\ref{fig5} as follows:
\begin{eqnarray}
 Y_d^{(2)} = (Y_e^{(2)})^T \sim
\frac{M_*}{\lambda^2 M_{\rm Pl}}
\left(
\begin{array}{ccc}
 \langle \xi_3 \rangle & \langle \xi_2 \rangle & \langle \xi_1 \rangle \\
\end{array}
\right)
\left(
\begin{array}{ccc}
 \lambda^6 & \lambda^5 & \lambda^3 \\
 \lambda^5 & \lambda^4 & \lambda^2 \\
 \lambda^3 & \lambda^2 & 1         \\
\end{array}
\right)
\left(
\begin{array}{c}
 \langle \xi_3 \rangle^T \\
 \langle \xi_2 \rangle^T \\
 \langle \xi_1 \rangle^T \\
\end{array}
\right)/M_*^2
\ .
\end{eqnarray}
Then,
the Yukawa couplings
for the down-type quarks and 
the charged leptons
are given by
\begin{eqnarray}
 Y_d &=& Y_e^T \sim Y_d^{(1)} + Y_d^{(2)} \nonumber \\
&\sim& \frac{M_*}{\lambda^3 M_{\rm Pl}}
\left(
\begin{array}{ccc}
 \langle \xi_3 \rangle & \langle \xi_2 \rangle & \langle \xi_1 \rangle \\
\end{array}
\right)
\left(
\begin{array}{ccc}
 \lambda^5 & \lambda^4 & \lambda^4 \\
 \lambda^4 & \lambda^3 & \lambda^3 \\
 \lambda^2 & \lambda   & \lambda   \\
\end{array}
\right)
\left(
\begin{array}{c}
 \langle \xi_3 \rangle^T \\
 \langle \xi_2 \rangle^T \\
 \langle \xi_1 \rangle^T \\
\end{array}
\right)/M_*^2\ .
\end{eqnarray}
This matrix has the same structure as
the ${\rm SU(5)_{GUT}}$ case and
gives the correct mass ratios and 
mixings.
The difference is the pre-factor $M_*/(\lambda^3 M_{\rm Pl})$.
The prediction for $\tan \beta$ depends
on the FN scale $M_*$ as follows:
\begin{eqnarray}
 \tan \beta \sim 16 \times 
\left( \frac{M_*}{M_{\rm GUT}} \right) \ .
\end{eqnarray}

\begin{figure}
\psfrag{16}{$16$}
\psfrag{xi2f}{$\lambda^3 \xi_1 + \lambda \xi_2 + \lambda^2 \xi_3$}
\psfrag{xi}{$\xi_1 + \lambda^2 \xi_2 + \lambda^3 \xi_3$}
\psfrag{S}{$S$}
\psfrag{Sb}{$\bar{S}$}
\psfrag{J}{$J$}
\psfrag{H10}{$H_{10}$}
\psfrag{lam3}{$\lambda^3 M_{\rm Pl}$}
\begin{center}
\includegraphics[width=10cm]{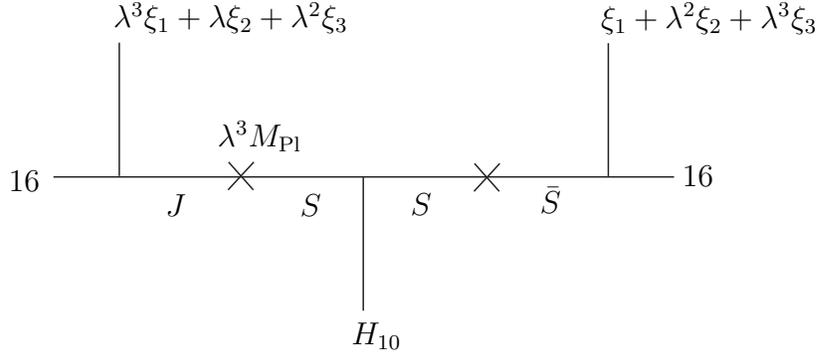} 
\end{center}
\caption{The Feynman diagram for the neutrino Dirac masses
in the SO(10) model.}
\label{fig6}
\end{figure}

The neutrino masses arise from a complicated mechanism.
First, the 9 $\times$ 9 Majorana mass 
matrix for the SU(5) singlet components
of $S$ ($1_S$), $\bar{S}$ ($1_{\bar{S}}$), and $16$ ($1_{16}$) 
are given by
\begin{eqnarray}
 M_{\rm Maj.} =
\bordermatrix{
             &    1_S                    & 1_{\bar{S}} & 1_{16} \cr
1_S          &  \lambda^{10} M_{\rm Pl}  &  M_*        & 0      \cr
1_{\bar{S}}  &  M_*  &  \lambda^{10} M_{\rm Pl} &  M_\xi^T      \cr
1_{16}       &  0    &  M_\xi            &  0                   \cr }
\ ,
\label{45}
\end{eqnarray}
where the submatrix $M_\xi$ is the Dirac mass terms
from the first line of eq.(\ref{37}):
\begin{eqnarray}
 M_\xi \sim M_*
\left(
\begin{array}{ccc}
 \lambda^3 & \lambda^3 & \lambda^3 \\
 \lambda^2 & \lambda^2 & \lambda^2 \\
 1         & 1         & 1 \\
\end{array}
\right)\ .
\end{eqnarray}
The ($1_{16}$, $1_{16}$) components of eq.(\ref{45})
are not exactly zero but are negligible
( $ \sim \lambda^{10} M_*^2/M_{\rm Pl} $ ).
The Majorana masses for the FN fields 
$1_S$ and $1_{\bar{S}}$ are of the order of $M_*$
which is too large for the seesaw mechanism.
However,
the suitable Majorana masses for the 
$1_{16}$ fields are given by
\begin{eqnarray}
 M_R \sim 
\lambda^{10} M_{\rm Pl} \cdot
\frac{M_{\xi} M_{\xi}^T}{M_*^2} \sim
\lambda^{10} M_{\rm Pl}
\left(
\begin{array}{ccc}
 \lambda^6 & \lambda^5 & \lambda^3 \\
 \lambda^5 & \lambda^4 & \lambda^2 \\
 \lambda^3 & \lambda^2 & 1 \\
\end{array}
\right)\ ,
\end{eqnarray}
The Dirac mass terms for the neutrinos 
are given by the Feynman diagram in Fig.\ref{fig6}.
From this diagram,
the structure of the Dirac mass is as follows:
\begin{eqnarray}
 m_{D} \sim \frac{M_* \langle H \rangle}{\lambda^2 M_{\rm Pl}}
\left(
\begin{array}{ccc}
 \langle \xi_3 \rangle &  \langle \xi_2 \rangle &  \langle \xi_1 \rangle \\
\end{array}
\right)
\left(
\begin{array}{ccc}
 \lambda & \lambda & \lambda \\
 1 & 1 & 1 \\
 \lambda^2 & \lambda^2 & \lambda^2 \\
\end{array}
\right)
\left(
\begin{array}{ccc}
 1 & 1 & 1 \\
 1 & 1 & 1 \\
 1 & 1 & 1 \\
\end{array}
\right) \nonumber \\
\times
\left(
\begin{array}{ccc}
 \lambda^3 & \lambda^2 & 1 \\
 \lambda^3 & \lambda^2 & 1 \\
 \lambda^3 & \lambda^2 & 1 \\
\end{array}
\right)/M_*
\ .
\end{eqnarray}
Then the neutrino masses originating from
the seesaw mechanism turn out to be
\begin{eqnarray}
 m_{\nu} &=& m_D M_R^{-1} m_D^T \nonumber \\
&\sim& \frac{M_*^2 \langle H \rangle^2}{\lambda^{14} M_{\rm Pl}^3}
\left(
\begin{array}{ccc}
 \langle \xi_3 \rangle &  \langle \xi_2 \rangle &  \langle \xi_1 \rangle \\
\end{array}
\right)
\left(
\begin{array}{ccc}
 \lambda & \lambda & \lambda \\
 1 & 1 & 1 \\
 \lambda^2 & \lambda^2 & \lambda^2 \\
\end{array}
\right)
\left(
\begin{array}{ccc}
 1 & 1 & 1 \\
 1 & 1 & 1 \\
 1 & 1 & 1 \\
\end{array}
\right)
\nonumber \\
&& \hspace*{6cm}
\times
\left(
\begin{array}{ccc}
 \lambda & 1 & \lambda^2 \\
 \lambda & 1 & \lambda^2 \\
 \lambda & 1 & \lambda^2 \\
\end{array}
\right)
\left(
\begin{array}{c}
 \langle \xi_3 \rangle^T \\
 \langle \xi_2 \rangle^T \\
 \langle \xi_1 \rangle^T \\
\end{array}
\right)/M_*^2 \nonumber \\
&\sim&
 \frac{M_*^2 \langle H \rangle^2}{\lambda^{14} M_{\rm Pl}^3}
\left(
\begin{array}{ccc}
 \langle \xi_3 \rangle &  \langle \xi_2 \rangle &  \langle \xi_1 \rangle \\
\end{array}
\right)
\left(
\begin{array}{ccc}
 \lambda^2 & \lambda & \lambda^3 \\
 \lambda   & 1       & \lambda^2 \\
 \lambda^3 & \lambda^2 & \lambda^4 \\
\end{array}
\right)
\left(
\begin{array}{c}
 \langle \xi_3 \rangle^T \\
 \langle \xi_2 \rangle^T \\
 \langle \xi_1 \rangle^T \\
\end{array}
\right)/M_*^2
\ ,
\label{49}
\end{eqnarray}
where the matrix 
whose components are all unity
in the second line
represents the 
following calculation as
\begin{eqnarray}
\left(
\begin{array}{ccc}
 \lambda^3 & \lambda^2 & 1 \\
 \lambda^3 & \lambda^2 & 1 \\
 \lambda^3 & \lambda^2 & 1 \\
\end{array}
\right)
\left(
\begin{array}{ccc}
 \lambda^6 & \lambda^5 & \lambda^3 \\
 \lambda^5 & \lambda^4 & \lambda^2 \\
 \lambda^3 & \lambda^2 & 1 \\
\end{array}
\right)^{-1}
\left(
\begin{array}{ccc}
 \lambda^3 & \lambda^3 & \lambda^3 \\
 \lambda^2 & \lambda^2 & \lambda^2 \\
 1 & 1 & 1 \\
\end{array}
\right)
\sim
\left(
\begin{array}{ccc}
 1 & 1 & 1 \\
 1 & 1 & 1 \\
 1 & 1 & 1 \\
\end{array}
\right)\ .
\end{eqnarray}
The matrix $m_\nu$ in eq.(\ref{49}) also has the same structure as 
the ${\rm SU(5)_{GUT}}$ model (eq.(\ref{24}))
which creates the suitable mass ratio and bi-maximal mixing
for the large angle MSW solution.
The pre-factor gives the appropriate
order of magnitude for the masses.

The results of the mixing and mass ratio
have been confirmed by
the numerical analysis in which
we diagonalize the whole mass matrix including 
FN fields.

\section{Conclusions and Discussions}

We have considered a horizontal ${\rm SU(3)_H \times U(1)_H}$ symmetry
to construct the appropriate mass matrices for the quarks and leptons.
We concentrated on the large mixing angle MSW solution for 
the solar neutrino problem.
The large mixing angle for $\nu_\mu \leftrightarrow \nu_\tau$
can be explained by the lopsided family structure.
In such a building block,
many models in the context of grand unified model 
predict that
the ratio of mass squared differences 
${\Delta m^2_{\rm solar}}/{\Delta m^2_{\rm atm}}$ 
is similar to $(m_c/m_t)^4$ \cite{Nomura:1999gm,Bludman:1992za}; 
otherwise, fine-tuning is needed to explain the hierarchy.

In this paper,
we have constructed realistic models in the context of
grand unified models
to explain the hierarchy naturally such that
\begin{equation}
\frac{\Delta m^2_{\rm solar}}{\Delta m^2_{\rm atm}} \sim 10^{-2}
\end{equation}
We emphasize that the unification of the three flavors enables us
to reproduce a neutrino mass matrix consistent with the 
large angle MSW solution.

Horizontal unification provides interesting features in
particle physics.
In supersymmetric models, flavor unification is a motivation
for the suppression of flavor changing neutral currents (FCNC)
\cite{Ellis:1982ts,Nir:1993mx}.
However, it is well known that the simple models 
with a gauged horizontal symmetry 
does not suppress the FCNC 
in the gravity-mediated supersymmetry 
breaking scenario \cite{Dine:1993np}.

It is interesting that the strong CP problem can be solved in our models.
The VEVs of the $\xi$ fields break 
the horizontal SU(3)$_{\rm H}$ symmetry.
The $\xi$'s have a Peccei-Quinn-like charge and 
the Peccei-Quinn mechanism 
will work well.
The invisible axion will be created if
the breaking scale $M_*$ is of the order of $10^{12}$ GeV.
The assumption of the horizontal scale ($M_* \sim 10^{12}$ GeV)
does not lead to gauge coupling divergence
below the GUT scale,
which is caused by 
the existence of the vector-like FN fields at the scale $M_*$.
Another attractive scale for $M_*$ is the GUT scale.
In the SO(10) unified model, the scale $M_*$ is constrained to be
$10^{15}$--$10^{16}$ GeV.
If the scale $M_*$ is at the GUT scale, too many axions will be created
and our universe will be overclosed. In such a case, 
we need the axions to be diluted by the inflation of the universe.

In our formulation, the horizontal unification
of SO(3)$_{\rm H}$ instead of SU(3)$_{\rm H}$ may be also available.
The SU(3)$_{\rm H}$ unification has the particularly interesting feature
that the representation of SU(3) is complex and 
spontaneous CP violation can occur \cite{Lee:1973iz}. 
The origin of the KM phase 
might be in the fundamental representation of $\xi$ in our model.

\section*{Acknowledgments}
This work was supported by JSPS Research Fellowships for 
Young Scientists (Y.M.).
Y.M. would like to thank to Prof. T. Yanagida for the useful suggestion
of the horizontal symmetries.
The authors would like to thank 
A. Akeroyd for reading the manuscript, and
J. Hisano, Y. Okada, and Y. Nomura for comments.


\begin{thebibliography}{99}
\newcommand{\wwwspires}{http://www.yukawa.kyoto-u.ac.jp/spires/find/hep/www}
%
%
\bibitem{CKM}
N. Cabibbo, Phys. Rev. Lett. {\bf 10}, 531 (1964); \\
M. Kobayashi and T. Maskawa, Prog. Theor. Phys. {\bf 49}, 652 (1973).

%
%
\bibitem{Groom:2000in}
D.~E.~Groom {\it et al.},
Eur.\ Phys.\ J.\  {\bf C15}, 1 (2000).

%
%
\bibitem{Fukuda:1998mi}
Y.~Fukuda {\it et al.}  [Super-Kamiokande Collaboration],
Phys.\ Rev.\ Lett.\  {\bf 81}, 1562 (1998).

%
%
\bibitem{Scholberg:1999ar}
K.~Scholberg  [SuperKamiokande Collaboration],
hep-ex/9905016.

%
%
%
\bibitem{Wolfenstein:1978ue}
L.~Wolfenstein,
Phys.\ Rev.\ {\bf D17}, 2369 (1978);\\
S.P.~Mikheev and A.Y.~Smirnov,
Sov.\ J.\ Nucl.\ Phys.\ {\bf 42}, 913 (1985);
Nuovo Cim.\ {\bf 9C}, 17 (1986).

%
%
\bibitem{ysuzuki}
Y.~Suzuki [SuperKamiokande Collaboration],
a talk given in the 19th International Conference
on Neutrino Physics and Astrophysics, Sudbury, Canada, (2000).

%
%
\bibitem{Apollonio:1999ae}
M.~Apollonio {\it et al.},
Phys.\ Lett.\  {\bf B466}, 415 (1999).

%
%
%
\bibitem{Froggatt:1979nt}
C.~D.~Froggatt and H.~B.~Nielsen,
Nucl.\ Phys.\  {\bf B147}, 277 (1979).

\bibitem{Leurer:1994gy}
M.~Leurer, Y.~Nir and N.~Seiberg,
Nucl.\ Phys.\  {\bf B420}, 468 (1994); 
L.~Ibanez and G.~G.~Ross,
Phys.\ Lett.\  {\bf B332}, 100 (1994); 
Y.~Grossman and Y.~Nir,
Nucl.\ Phys.\  {\bf B448}, 30 (1995); 
E.~Dudas, S.~Pokorski and C.~A.~Savoy,
Phys.\ Lett.\  {\bf B356}, 45 (1995); 
P.~Binetruy, S.~Lavignac and P.~Ramond,
Nucl.\ Phys.\  {\bf B477}, 353 (1996); 
P.~Binetruy, S.~Lavignac, S.~Petcov and P.~Ramond,
Nucl.\ Phys.\  {\bf B496}, 3 (1997); 
A.~E.~Nelson and D.~Wright,
Phys.\ Rev.\  {\bf D56}, 1598 (1997); 
N.~Irges, S.~Lavignac and P.~Ramond,
Phys.\ Rev.\  {\bf D58}, 035003 (1998); 
J.~K.~Elwood, N.~Irges and P.~Ramond,
Phys.\ Rev.\ Lett.\  {\bf 81}, 5064 (1998); 
Y.~Grossman, Y.~Nir and Y.~Shadmi,
JHEP {\bf 9810}, 007 (1998); 
G.~Altarelli and F.~Feruglio,
JHEP {\bf 9811}, 021 (1998); 
K.~Choi, K.~Hwang and E.~J.~Chun,
Phys.\ Rev.\  {\bf D60}, 031301 (1999); 
Q.~Shafi and Z.~Tavartkiladze,
Phys.\ Lett.\  {\bf B451}, 129 (1999);
M.~Bando and T.~Kugo,
Prog.\ Theor.\ Phys.\  {\bf 101}, 1313 (1999);
S.~Lola and G.~G.~Ross,
Nucl.\ Phys.\  {\bf B553}, 81 (1999); 
Y.~Nir and Y.~Shadmi,
JHEP {\bf 9905}, 023 (1999); 
S.~F.~King,
Nucl.\ Phys.\  {\bf B562}, 57 (1999); 
Q.~Shafi and Z.~Tavartkiladze,
Phys.\ Lett.\  {\bf B487}, 145 (2000);
S.~F.~King,
Nucl.\ Phys.\  {\bf B576}, 85 (2000); 
J.~Ellis, G.~K.~Leontaris and J.~Rizos,
JHEP {\bf 0005}, 001 (2000).

%
%
\bibitem{Wilczek:1979xi}
F.~Wilczek and A.~Zee,
Phys.\ Rev.\ Lett.\  {\bf 42}, 421 (1979);
Z.~G.~Berezhiani and D.~L.~Chkareuli,
Sov.\ J.\ Nucl.\ Phys.\  {\bf 37}, 618 (1983);
R.~Barbieri, L.~J.~Hall, S.~Raby and A.~Romanino,
Nucl.\ Phys.\  {\bf B493}, 3 (1997);
G.~Eyal,
Phys.\ Lett.\  {\bf B441}, 191 (1998);
T.~Blazek, S.~Raby and K.~Tobe,
Phys.\ Rev.\  {\bf D60}, 113001 (1999);
M.~Tanimoto, T.~Watari and T.~Yanagida,
Phys.\ Lett.\  {\bf B461}, 345 (1999);
R.~Dermisek and S.~Raby,
Phys.\ Rev.\  {\bf D62}, 015007 (2000); 
T.~Blazek, S.~Raby and K.~Tobe,
Phys.\ Rev.\  {\bf D62}, 055001 (2000). 

%
%
\bibitem{Fritzsch}
H. Fritzsch, Phys. Lett. {\bf 73B}, 317 (1978);
Nucl. Phys. {\bf B115}, 189 (1979).

%
%
\bibitem{Branco}
G.C. Branco and J.I. Silva-Macros, Phys. Lett. {\bf B331}, 390 (1994).

%
%
%
\bibitem{Sato:1998hv}
J.~Sato and T.~Yanagida,
Phys.\ Lett.\  {\bf B430}, 127 (1998);\\
T.~Yanagida and J.~Sato,
Nucl.\ Phys.\ Proc.\ Suppl.\  {\bf 77}, 293 (1999).

%
%
\bibitem{Albright:1999jv}
C.~H.~Albright and S.~M.~Barr,
Phys.\ Lett.\  {\bf B452}, 287 (1999).


\bibitem{GUT}
E. Witten, Nucl. Phys. {\bf B188}, 513 (1981); \\
S. Dimopoulos, S. Raby, and F. Wilczek, Phys. Rev. {\bf D24}, 1861 (1981); \\
S. Dimopoulos and H. Georgi, Nucl. Phys. {\bf B193}, 150 (1981); \\
N. Sakai, Zeit. Phys. {\bf C11}, 153 (1981).

%
%
\bibitem{seesaw}
T.~Yanagida,
in
{\em Proceedings of the Workshop on Unified Theory
and Baryon Number of the Universe},
eds. O. Sawada and A. Sugamoto (KEK, 1979) p.95;\\
M.~Gell-Mann, P.~Ramond, and R.~Slansky,
in {\em Supergravity},
eds. P.~van Nieuwenhuizen and D.~Freedman
(North Holland, Amsterdam, 1979).

\bibitem{Fusaoka}
H.~Fusaoka and Y.~Koide,
Phys.\ Rev.\  {\bf D57}, 3986 (1998).

%
%
%
\bibitem{Maki:1962mu}
Z.~Maki, M.~Nakagawa and S.~Sakata,
Prog.\ Theor.\ Phys.\  {\bf 28}, 870 (1962).

%
%
\bibitem{Harvey}
J.A. Harvey, D.B. Reiss, and P. Ramond,
Nucl. Phys. {\bf B199}, 223 (1982).

%
%
%
\bibitem{Wilczek:1982rv}
F.~Wilczek,
Phys.\ Rev.\ Lett.\  {\bf 49}, 1549 (1982); \\
G.~B.~Gelmini, S.~Nussinov and T.~Yanagida,
Nucl.\ Phys.\  {\bf B219}, 31 (1983); \\
J.~L.~Feng, T.~Moroi, H.~Murayama and E.~Schnapka,
Phys.\ Rev.\  {\bf D57}, 5875 (1998).

%
%
%
\bibitem{Peccei:1977hh}
R.~D.~Peccei and H.~R.~Quinn,
Phys.\ Rev.\ Lett.\  {\bf 38}, 1440 (1977); 
Phys.\ Rev.\  {\bf D16}, 1791 (1977).

%
%
\bibitem{Kim:1979if}
J.~E.~Kim,
Phys.\ Rev.\ Lett.\  {\bf 43}, 103 (1979); \\
M.~A.~Shifman, A.~I.~Vainshtein and V.~I.~Zakharov,
Nucl.\ Phys.\  {\bf B166}, 493 (1980).

%
%
\bibitem{Zhitnitsky:1980tq}
A.~R.~Zhitnitsky,
Sov.\ J.\ Nucl.\ Phys.\  {\bf 31} (1980) 260; \\
M.~Dine and W.~Fischler,
Phys.\ Lett.\  {\bf B120}, 137 (1983).

%
%
%
\bibitem{Guth:1981zm}
A.~H.~Guth,
Phys.\ Rev.\  {\bf D23}, 347 (1981).

%
%
%
\bibitem{Albright:1998vf}
C.~H.~Albright, K.~S.~Babu and S.~M.~Barr,
Phys.\ Rev.\ Lett.\  {\bf 81}, 1167 (1998); \\
K.~S.~Babu, J.~C.~Pati and F.~Wilczek,
Nucl.\ Phys.\  {\bf B566}, 33 (2000); \\
Y.~Nomura and T.~Sugimoto,
Phys.\ Rev.\  {\bf D61}, 093003 (2000).

%
%
\bibitem{Nomura:1999gm}
Y.~Nomura and T.~Yanagida,
Phys.\ Rev.\  {\bf D59}, 017303 (1999).

%
%
%
\bibitem{Lucas:1997bc}
V.~Lucas and S.~Raby,
Phys.\ Rev.\  {\bf D55}, 6986 (1997); \\
T.~Goto and T.~Nihei,
Phys.\ Rev.\  {\bf D59}, 115009 (1999).

%
%
\bibitem{Bludman:1992za}
S.~A.~Bludman, D.~C.~Kennedy and P.~G.~Langacker,
Phys.\ Rev.\  {\bf D45}, 1810 (1992); \\
V.~Barger, S.~Pakvasa, T.~J.~Weiler and K.~Whisnant,
Phys.\ Lett.\  {\bf B437}, 107 (1998); \\
C.~H.~Albright and S.~M.~Barr,
Phys.\ Lett.\  {\bf B461}, 218 (1999).

%
%
\bibitem{Ellis:1982ts}
J.~Ellis and D.~V.~Nanopoulos,
Phys.\ Lett.\  {\bf B110}, 44 (1982).

%
%
\bibitem{Nir:1993mx}
Y.~Nir and N.~Seiberg,
Phys.\ Lett.\  {\bf B309}, 337 (1993); \\
P.~Pouliot and N.~Seiberg,
Phys.\ Lett.\  {\bf B318}, 169 (1993); \\
R.~Barbieri, G.~Dvali and L.~J.~Hall,
Phys.\ Lett.\  {\bf B377}, 76 (1996).

%
%
%
\bibitem{Dine:1993np}
M.~Dine, R.~Leigh and A.~Kagan,
Phys.\ Rev.\  {\bf D48}, 4269 (1993); \\
K.~S.~Babu and R.~N.~Mohapatra,
Phys.\ Rev.\ Lett.\  {\bf 83}, 2522 (1999).

%
%
%
%
\bibitem{Lee:1973iz}
T.~D.~Lee,
Phys.\ Rev.\  {\bf D8}, 1226 (1973); \\
S.~Weinberg,
Phys.\ Rev.\ Lett.\  {\bf 37}, 657 (1976); \\
G.~C.~Branco,
Phys.\ Rev.\  {\bf D22}, 2901 (1980).

\end{thebibliography}
\end{document}